\documentclass[final,3p,times]{elsarticle}

\usepackage{epsfig,amssymb,amsmath,listings,lineno,multirow,rotating,amsfonts,colortbl,color,soul}
\usepackage[utf8]{inputenc}

\newcommand{\bl}[1]{\boldsymbol{#1}}
\newcommand{\p}{_{\rm p}}
\newcommand{\f}{_{\rm f}}
\renewcommand{\r}{_{\rm r}}
\renewcommand{\c}{_{\rm c}}
\newcommand{\g}{\cellcolor[gray]{0.9}}
\begin{document}
\begin{frontmatter}
\title{A correction scheme for two-way coupled point-particle simulations on anisotropic grids}

\author[cornell,CTR]{M. Esmaily\corref{cor1}}
\ead{me399@cornell.edu}
\author[stanford]{J. A. K. Horwitz}
\ead{horwitz1@stanford.edu}
\address[cornell]{Sibley School of Mechanical and Aerospace Engineering, Cornell University, Ithaca, NY 14850, USA}
\address[CTR]{Center for Turbulence Research, Stanford University, Stanford, CA 94305, USA}
\address[stanford]{Department of Mechanical Engineering, Stanford University, Stanford, CA 94305, USA}
\cortext[cor1]{Corresponding author}

\begin{abstract}
The accuracy of Lagrangian point-particle models for simulation of particle-laden flows may degrade when the particle and fluid momentum equations are two-way coupled. 
In these cases the fluid velocity at the location of the particle, which is often used as an estimation of the undisturbed velocity, is altered by the presence of the particle, modifying the slip velocity and producing an erroneous prediction of coupling forces between fluid and particle. 
In this article, we propose a correction scheme to eliminate this error and predict the undisturbed fluid velocity accurately. 
Conceptually, in this method, the computation cell is treated as a solid object immersed in the fluid that is subjected to the two-way coupling force and dragged at a velocity that is identical to the disturbance created by the particle. 
The proposed scheme is generic as it can be applied to unstructured grids with arbitrary geometry and particles that have different size and density. 
At its crudest form for isotropic grids, the present correction scheme reduces to dividing the Stokes drag by $1 - 0.75\Lambda$, where $\Lambda$ is the ratio of the particle diameter to the grid size.
The accuracy of the proposed scheme is evaluated by comparing the computed settling velocity of individual and pair of particles under gravity on anisotropic rectilinear grids against analytical solutions.
This comparison shows up to two orders of magnitude reduction in error in cases where the particle is up to 5 times larger than the grid that may have an aspect ratio of over 10.
Furthermore, a comparison against the particle-resolved simulation of decaying turbulence demonstrates the excellent accuracy of the proposed scheme.
\end{abstract}
\end{frontmatter}

\section{Introduction}
Flows laden with particles are commonplace in nature, ranging from the transport of droplets in clouds \citep{shaw2003} and pollution in environmental flows \cite{zhang2004evolution}, to spray combustion \cite{pitsch2008large} and particle-based solar receivers in engineering related flows \citep{farbar2016monte, pouransari2017effects}. 
Modeling such flows are of importance for better understanding of underlying physics, making better predictions, and designing better devices. 
Among the wide verity of modeling approaches, the Eulerian-Lagrangian point-particle method has been an attractive choice for its favorable accuracy, simplicity, and affordability \cite{vie-2014, elghobashi1994predicting, balachandar2009scaling}. 
In this method, the flow is modeled using a conventional Eulerian framework, and the individual particles are treated as points with a finite mass that experience a drag force, which is calculated using a reduced-order model.
Although the point-particle approach offers advantages over other techniques, e.g., allows for particle trajectory crossing in comparison to a fully Eulerian technique, its accuracy can deteriorate in certain scenarios. 
One scenario, which is of particular interest in this study, is when the fluid and particle momentum equations are two-way coupled, i.e., a force opposite of the force that is exerted by the fluid to the particle is applied to the fluid according to Newton's third law. 
The two-way coupling force changes the velocity of the fluid at the location of the particle, modifying it to a quantity different from the undisturbed fluid velocity \cite{garg2007accurate}. 

To remedy this issue and reduce the error in two-way coupled point-particle calculations, a number of methods have been proposed \citep{horwitz2016accurate, gualtieri_etal_2015, ireland_desjardins_2016}. 
In one approach described in \citep{horwitz2016accurate}, the disturbance created by the presence of the particle is estimated based on the enhanced curvature in the fluid and employed to correctly model the Stokes drag exerted on the particle.
In a different approach proposed in \citep{gualtieri_etal_2015, ireland_desjardins_2016}, the disturbance is estimated according to the analytical solution of a Gaussian-regularized point-force that is distributed over a volume which scales with the particle size. 
These studies have shown that the two-way coupling error is primarily a function of the ratio of the grid size to the particle size, becoming very large for grids smaller than the particle. 
The former scheme was developed as an efficient procedure for isotropic grids, while the latter procedures, though applicable for anisotropic grids, require considerable computational resources owing to the number of points over which the Gaussian stencil must be resolved. 
These conditions limit these schemes' utility for the combined goal of accuracy and efficiency of two-way coupled computation in arbitrary grid configurations. 

The motivation of present study is to devise a procedure that is general in scope and efficient in terms of computational cost. 
Despite the generality of our formulation and its applicability to any unstructured grid, particular attention is given to rectilinear grids for its application in studying wall-bounded particle-laden turbulent flows \citep{rogers_eaton_1991}.
These flows are typically simulated on grids with aspect ratios over 10, requiring a correction procedure that handles highly-skewed cells which can become smaller than the particle \cite{segura2004} near solid boundaries. 
Therefore, the target of this study is to develop an efficient correction scheme that can be utilized for point-particle simulations on a wide range of grid geometries and sizes with a reasonable accuracy.

To establish the accuracy of the proposed scheme we rely on analytical predictions and particle-resolved simulations as the reference solutions.
In total, we consider four sets of test cases. 
In the first set, following the procedure in \citep{horwitz2016accurate}, we consider settling of a single particle using an isotropic grid. 
The extension of the first set to anisotropic grids will be considered in the second set of test cases.
In the third set, the settling velocity of two nearby particles is compared against an analytical solution to investigate the scheme's accuracy in accounting for particle-particle screening. 
As the final case, we simulate a decaying turbulence laden with inertial particles and compare our results against a particle-resolved simulation \cite{subramaniam_et_al_2014} to examine the accuracy of our scheme in a more realistic setting, where the undisturbed fluid velocity owing to the temporal and spatial variation of the flow is not well defined.
A simplified version of this correction scheme, which can be implemented in practice by adding a prefactor to the Stokes drag equation, is described in \ref{sec:simplified}. 






\section{A correction scheme} \label{sec:method}
In what follows, we use Roman subscript to distinguish various parameters (e.g. subscript f in $\bl u\f$ refers to fluid and p in $\bl u\p$ refers to particle), italic subscripts to denote computational grid points (e.g., $\bl F_j$ denotes force exerted to point $j$), and parenthesize italic superscripts to denote directions in an Euclidean 3-dimensional space (e.g. $a^{(1)}$ denotes the grid size in direction 1).
In this section, we first introduce the underlying concept and then construct the correction scheme step-by-step using a physics-based approach. 

\subsection{The concept}
Consider the equation of motion of a particle expressed as
\begin{equation}
m\p \dot {\bl u}\p = \bl F + m\p \bl g,
\label{particle_eq}
\end{equation}
where $m_p$ is the mass of the particle, $\bl g$ is the external force applied to the particle, $\bl u\p$ is the particle velocity, $\dot{(\bullet)}$ denotes Lagrangian time derivative, and $\bl F$ accounts for the two-way coupling forces that are exerted from the flow to the particle. 
In the point-particle approach $\bl F$ is typically modeled as
\begin{equation}
\bl F = \bl F_{\rm d} + \bl F_{\rm b} + \bl F_{\rm l} + \bl F_{\rm h} + \bl F_{\rm a} \cdots,
\label{F_exp}
\end{equation}
where $\bl F_{\rm d}$ denoting the drag force in a steady-state flow, $\bl F_{\rm b}$ is the buoyancy force, $\bl F_{\rm l}$ is the lift force induced in flows with strong shear, $\bl F_{\rm h}$ is the force created by the history effect, and $\bl F_{\rm a}$ is the added mass force. 
Depending on the regime under consideration, other terms can appear in Eq. \eqref{F_exp} to model for instance the influence of thermal fluctuation on particle motion. 
We only include $\bl F_{\rm d}$ and $\bl F_{\rm b}$ in our case studies.
The following formulation, nevertheless, is based on a generic $\bl F$ and can be applied to cases where other forces are present. 

The existing reduced-order models for various elements of $\bl F$ in Eq. \eqref{F_exp} are reliant on the knowledge of the undisturbed velocity. 
The undisturbed velocity is by definition unaffected by the presence of the particle. 
Taking for instance $\bl F_{\rm d}$ and assuming small particle Reynolds number, one can express 
\begin{equation}
\bl F_{\rm d} = 3\pi \mu d\p (\bl u\f - \bl u\p).
\label{stokes}
\end{equation}
where $d\p$ is the particle diameter, $\mu$ is the fluid dynamic viscosity, and $\bl u\f$ is the undisturbed velocity.

The definition of $\bl u\f$ stems from the analytical derivation of Eq. \eqref{stokes}, in which the flow is uniform in the far-field of a single particle and $\bl u\f$ is a constant.
In practical applications, where the flow is varying arbitrarily in space with arbitrarily small structures, $\bl u\f$ does not have a well-defined meaning. 
It is common in computational fluid dynamics codes to take the fluid velocity at the location of the particle as the undisturbed velocity. 
The shortcoming of this approach is that the fluid velocity at the location of the particle is modified (disturbed) by the presence of a particle. 
As a result, the prediction of the Stokes drag model in Eq. \eqref{stokes} is altered by the wrong estimation of $\bl u\f$, producing a wrong $\bl F_{\rm d}$, and consequently inaccurate particle and fluid statistics \citep{MehrabadiJFM}.
The degree to which the estimation of $\bl u\f$ deviates from its actual value depends on some factors such as the size of the particle relative to the computational grid, the interpolation scheme, and the particle Stokes and Reynolds numbers \cite{horwitz2016accurate}.
The aim of this article is to devise a procedure for accurate computation of $\bl u\f$.
More specifically, this procedure is designed for the Lagrangian point-particle approach embedded in a Eulerian fluid solver where the influence of the particle on the fluid is captured through a point force exerted at the location of the particle.

To estimate $\bl u\f$, we begin by attempting to answer the following question: 
Suppose we have a computational domain, where the velocity is represented by a set of discrete points.
In general, different components of the velocity may correspond to different points in space, as is the case with a staggered grid. 
Additionally, suppose the fluid is completely at rest in this domain at $t=0$.
Then, we apply an infinitesimal force $\bl F$ to one of these computational grid points. 
In the discrete setting, this operation translates to perturbing one of the entries of the discretized forcing on the right-hand side of the Navier-Stokes equation. 
After applying this force, the fluid accelerates till reaching a steady state at $t \to \infty$. 
Then we measure the fluid velocity at that computational grid point and denote it by the velocity of the computational cell $\bl u\c$.
The main question is, what is $\bl u\c$ as a function of fluid properties, the grid dimensions, and $\bl F$?

The answer to the above question lies at the foundation of our correction scheme.
By devising a scheme to compute $\bl u\c$,  we can subtract it from the disturbed fluid velocity, denoted by $\bl u_{\rm d}$, to obtain
\begin{equation}
\bl u\f = \bl u_{\rm d} - \bl u\c,
\label{ufe_def}
\end{equation}
which is an estimation of the undisturbed fluid velocity.
To demonstrate how Eq. \eqref{ufe_def} operates with an example, consider the above scenario where the fluid is quiescent, and the undisturbed velocity is zero. 
Thus, a perfect correction scheme should produce $\bl u\f = 0$.
Since by definition the disturbed velocity $\bl u_{\rm d}$ is identical to $\bl u\c$, Eq. \eqref{ufe_def} indeed predicts $\bl u\f = 0$.
Therefore, Eq. \eqref{ufe_def} can be used as the basis of a correction scheme once $\bl u\c$ is known.

What is left to be done is to devise a method that yields an estimate of $\bl u\c$, given the fluid and grid properties as well as $\bl F$. 
As stated above, $\bl u\c$ is generated as a result of a point-force exerted to the computational cell.
In the discrete setting, $\bl u\c$ represents the velocity of a finite volume of the fluid that its size scales with the computational cell. 
To compute $\bl u\c$, we need to estimate the response of that volume of the fluid to a point force. 
Provided that the flow within that volume does not vary spatially and is represented by a single velocity $\bl u\c$, we model the computational cell as a solid object \footnote{In this model, we neglect higher order variation of the velocity within the computational cell that translates to solid-body rotation of the solid object and its bending.
The higher order effects are of secondary importance in computational $\bl u\c$.}.
Therefore, to estimate $\bl u\c$, we solve for the velocity of a small solid object with dimensions similar to the computational cell that is surrounded by the fluid and subjected to a point force $\bl F$. 

For the equation of the motion of a small solid object in a fluid, we consider a generalized Maxey-Riley equation \cite{maxey1983equation}. 
Since $\bl u\c$ is the deviation of the cell velocity from a base undisturbed flow, the undisturbed velocity of that solid object itself is zero. 
Hence, the equation of the motion of that solid object (which hereafter is denoted by the computational cell) in direction $i$ simplifies to
\begin{equation}
\frac{3}{2}m\c \dot u\c^{(i)} = -3\pi\mu d\c K_{\rm t}^{(i)} u\c^{(i)} - F^{(i)},
\label{uc_ge}
\end{equation}
where $d\c$ is the equivalent diameter of the computational cell, $m\c = (\pi/6) \rho\f d\c^3$ is the mass of the computational cell, $K_{\rm t}^{(i)}$ is the drag correction factor defined below, and $\nu=\mu/\rho\f$ is the kinematic viscosity of the fluid.
In cases where the particle is much larger than the size of the computational cell, one may choose to use the particle-displaced mass of the fluid rather than the mass of fluid in the computational cell to compute $m\c$.
In our computations, we take $m\c = (\pi/6) \rho\f \max(d\c,d\p/2)^3$ that relieves the time-step size requirement of our time integration scheme when $d\p > 2d\c$ \footnote{$d\p/2$ rather than $d\p$ is employed, since it is sufficiently large to produce an unconditionally stable scheme.}. 
Apart from the stability implications, using a larger value for $m\c$ has a minimal effect on the accuracy of the scheme.
Note that the acceleration term is multiplied by $3/2$ in Eq. \eqref{uc_ge} to account for the added mass.
As discussed in \ref{app:history}, the history effects can be incorporated into Eq. \eqref{uc_ge}, particularly for regimes in which the transient effects are important.
However, given the added complexity and its minimal overall effect, we neglect the history effects in all the calculations and case studies that follow.

The equivalent diameter of the computational cell in Eq. \eqref{uc_ge} is defined as the diameter of a sphere that has the same volume as the computational cell $\mathbb V$, thus in general $\pi d\c^3 = 6 \mathbb{V}$.
Since we will be considering only rectilinear grids, the computational cell is a cuboid with $a^{(1)}$, $a^{(2)}$, and $a^{(3)}$ sides.
Thus, $d\c$ can be expressed as
\begin{equation}
d\c = \left( \frac{6}{\pi}a^{(1)}a^{(2)}a^{(3)}\right)^{1/3}.
\label{dc}
\end{equation}

Aside from the procedure for computation of $K_{\rm t}^{(i)}$, the present correction scheme is in essence integrating Eq. \eqref{uc_ge} to obtain $\bl u\c$ and using it to correct $\bl u_{\rm d}$ via Eq. \eqref{ufe_def}.
Given the two-way coupling force $\bl F$, Eq. \eqref{uc_ge} is integrated concurrent with the particle equation of motion (Eq. \eqref{particle_eq}) to obtain $\bl u\c$.
Thus, the computational cell velocity $\bl u\c$ is solved in a Lagrangian frame that is attached to the particle, viz. a single $\bl u\c$ is computed for each particle and used to obtain that particle's undisturbed velocity.
As outlined in \ref{sec:simplified}, the overall scheme can be further simplified to a closed-form algebraic equation if the time-dependent term in Eq. \eqref{uc_ge} is neglected.
$K_{\rm t}^{(i)}$ can also be substituted with a constant as a crude estimate.
In that simplified form, the present correction reduces to a pre-factor in the Stokes drag formula, which in the case of an isotropic grid is simply $(1 - 0.75 d\p/a)^{-1}$ (\ref{sec:simplified}).

The role of $K_{\rm t}^{(i)}$ is to make fine adjustment to the drag coefficient by capturing several effects.
It is expressed as
\begin{equation}
K_{\rm t}^{(i)} = \frac{ K\c^{(i)} C\r}{K\p^{(i)} C_{\rm t}^{(i)}}.
\label{Kt}
\end{equation}
In this equation, $K\c^{(i)}$, which only depends on the cell geometry, accounts for the non-sphericity of the computational cell.
$K\p^{(i)}$, which primarily depends on the interpolation scheme, accounts for the distribution of $\bl F$ among multiple computational cells.
$C\r$ is a function of the cell Reynolds number and accounts for the increase in drag at finite Reynolds numbers.
$C_{\rm t}^{(i)}$, which is a function of the cell relaxation time and the particle CFL number, accounts for the limited exposure of a particle to a computational cell.

Before elaborating on the role of each of these coefficients and describe a procedure for their calculation, we need to make a side note on the fundamental difference between external flows in three and two dimensions that prohibits one from extending the present correction to a 2-dimensional setting.
In three dimensions, the disturbance created by the particle vanishes at infinity since it is inversely proportional to the distance from the particle (one can simply show this by computing the Green's function of the Poisson's equation in three dimensions). 
Having a disturbance that vanishes at infinity allows one to relate the undisturbed velocity to the fluid velocity at the location of the particle such that the difference remains finite and the problem well-posed. 
In 2-dimensional creeping flow, however, the disturbance does not vanish at infinity due to the well known Stokes' paradox.
As a result, the correction discussed here should only be used in 3-dimensional settings. 

\subsection{The geometric correction factor}
\begin{figure}
\begin{center}
\includegraphics[width=0.4\textwidth]{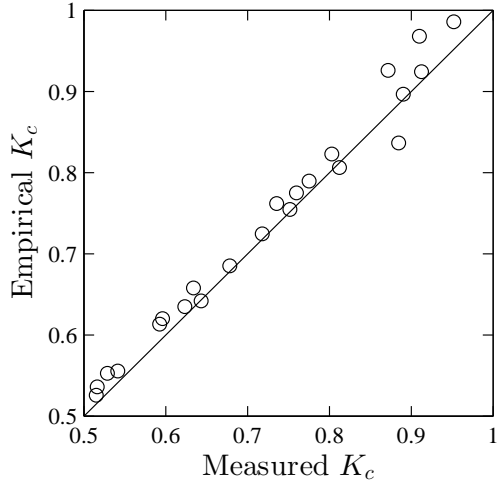}
\caption{The prediction of Eq. \eqref{kc} versus the numerical measurement of $K\c^{(i)}$ for cuboid cells that have aspect ratio ranging from 1 to 16.}
\label{fig:Kc}
\end{center}
\end{figure}

At small Reynolds numbers, contrary to a sphere, the drag on an object with an arbitrary geometry is not known analytically.
The geometrical correction factor $K\c^{(i)}$ must account for the change in the drag associated with the computational cell geometry when a sphere is considered to be the reference geometry.
For this purpose, we suggest using empirical relationships that are available in the literature for computing the drag coefficient of simple geometries at low Reynolds number.
In  \cite{leith1987drag}, several relationships are provided for prisms and in particular cuboid to express $K\c^{(i)}$ as a function of $a^{(j)}$'s.
To identify the most accurate expression, we performed several numerical experiments, in which we vary the cells aspect ratios $a^{(2)}/a^{(1)}$ and $a^{(3)}/a^{(1)}$ from 1 to 16 and measure $K\c^{(i)}$ directly.
In these computations, we exert a small force to a computational cell and measure the resulting velocity once conditions are steady.
From scaling analysis \citep{batchelor-1967}, it can be shown $\|\bl F\| \ll \nu\mu$ would satisfy the infinitesimal force requirement, a condition that translates to a small Reynolds number for the computational cell. 
Since the force is fixed in space, small, and exerted to one computational cell, $C_{\rm t}^{(i)}$, $C\r$, and $K\p^{(i)}$ are all equal to one, respectively.
Then, $K\c^{(i)}$ can be measured directly using Eq. \eqref{uc_ge}.
The best fit to data show
\begin{equation}
K\c^{(i)} = 1.52 - 0.83\left(\frac{d\c}{d_{\rm s}}\right)^2 -0.35\left(\frac{d\c}{d_{\rm n}^{(i)}}\right) + 0.056\left(\frac{\max\left(a^{(1)},a^{(2)},a^{(3)}\right)}{d_{\rm n}^{(i)}}\right),
\label{kc}
\end{equation}
is an empirical equation that provides the best prediction for $K\c^{(i)}$ (Figure \ref{fig:Kc}).
In Eq. \eqref{kc}, $K_c^{(i)}$ denotes the correction to the Stokes drag coefficient in the direction $i$ and
\begin{equation}
\begin{split}
d_{\rm s}       &= \sqrt{\frac{2}{\pi}(a^{(1)}a^{(2)}+a^{(2)}a^{(3)}+a^{(3)}a^{(1)})}, \\
d_{\rm n}^{(i)} &= \sqrt{\frac{4a^{(1)}a^{(2)}a^{(3)}}{\pi a^{(i)}}},
\end{split}
\label{ds_dn}
\end{equation}
are the diameter of the respective spheres that have the same surface area and frontal area as the computational cell, respectively.
Based on Eq. \eqref{kc}, $K\c^{(i)}=0.516$ for an isotropic grid with $a^{(1)}=a^{(2)}=a^{(3)}$.

\subsection{The interpolation correction factor}
Next, we derive an analytical relationship for $K\p^{(i)}$ for an interpolation scheme and a cell geometry that are arbitrary.
$K\p^{(i)}$ should account for the fact that $\bl F$ is not always applied to a particular computational cell but distributed among cells adjacent to the particle.
Conceptually, we accomplish this by finding the disturbance field around a given computational cell interpolating that to the location of the particle. 

The two-way coupling force $\bl F$ exerted to the fluid at the location of particle is distributed among the computational cells according to 
\begin{equation}
F_j^{(i)} = \beta_j^{(i)} F^{(i)},
\label{force_dist}
\end{equation}
where $\beta_j^{(i)}$ is the interpolation coefficient associated with the computational cell $j$ in direction $i$ and $\bl F_j$ is the force exerted to the computational cell $j$. 
Note that for cuboid cells and a trilinear interpolation scheme $j \in\{1,\cdots,8\}$.
Additionally, the computational cell velocity at the location of particle $\bl u\c$ is interpolated from the adjacent computational cells using
\begin{equation}
u\c^{(i)} = \gamma_j^{(i)} u_{{\rm c}, j}^{(i)},
\label{vel_int}
\end{equation}
where $\gamma_j^{(i)}$ and $u_{{\rm c},j}^{(i)}$ are the interpolation coefficient and the velocity associated with the computational cell $j$ in direction $i$.
In our framework, we use the same interpolation scheme for distributing $\bl F$ among computational cells and computing $\bl u\c$ at the particle position, thus $\gamma_j^{(i)} = \beta_j^{(i)}$.

\begin{figure}
\begin{center}
\includegraphics[width=0.3\textwidth]{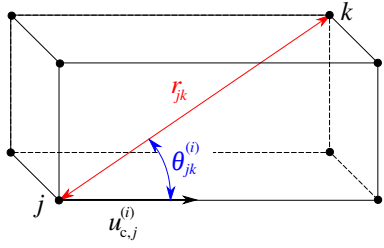}
\caption{Schematic of a computational cell $j$ with velocity $u_{{\rm c},j}^{(i)}$ in direction $i$ produced by applying an infinitesimal force to the cell $j$ in the direction $i$.
In relation to an adjacent computational cell $k$, $r_{jk}$ is the depicted length normalized by $0.28 d\c$ and $\theta_{jk}^{(i)}$ is the depicted polar angle. }
\label{fig:Kp-schematic}
\end{center}
\end{figure}

Next, let us denote the disturbance introduced at the computational cell $k$ when the computational cell $j$ is perturbed in direction $i$ by $\alpha_{jk}^{(i)}$. 
In other words, $\alpha_{jk}^{(i)}$ is the ratio between the velocity of the computational cell $k$ to that of $j$ when an infinitesimal force is applied to the computational cell $j$.
Our numerical experiments shows that $\alpha_{jk}^{(i)}$ can be estimated based on
\begin{equation}
\alpha_{jk}^{(i)} = \frac{3}{4} r_{jk}^{-1}\left(1 + \cos^2\theta_{jk}^{(i)} \right) + \frac{1}{4} r_{jk}^{-3}\left(1 - 3\cos^2 \theta_{jk}^{(i)} \right),
\label{alpha}
\end{equation}
which is the velocity field around a sphere at the limit of zero Reynolds number. 
At this limit, the flow around a sphere is symmetric as manifested in this expression by $\alpha_{jk}^{(i)}=\alpha_{kj}^{(i)}$.
At finite Reynolds numbers, a more elaborate formulation would be required for capturing the asymmetry of the flow as shown in Section \ref{sec:settling}.
In Eq. \eqref{alpha}, $r_{jk}$ is the distance between the computational cell $j$ and $k$ normalized by the characteristic length of the computational cell and $\theta_{jk}^{(i)}$ is the polar angle between the line passing through computational cells $j$ and $k$ and the direction of travel $i$ (Figure \ref{fig:Kp-schematic}).
For a flow around a sphere, $r_{jk}$ is normalized based on the radius of the sphere.
Our numerical results, however, show that $0.28 d\c$ (rather than $0.5 d\c$) is a better characteristic length by providing a better estimation of $\alpha_{jk}^{(i)}$.
If this normalization leads to $r_{jk} < 1$, which may occur for extremely skewed cells, we take $r_{jk} = 1$ instead.
The accuracy of Eq. \eqref{alpha} is shown in Table \ref{table:far_field}, where its prediction is compared against the measured $\alpha_{jk}^{(i)}$ that is obtained by measuring the velocity of the computational cells surrounding a perturbed computational cell. 

\begin{table}
\centering
\begin{tabular}{ccc|cccccccc}
  \hline \hline
   $a^{(2)}/a^{(1)}$ & $a^{(3)}/a^{(1)}$ &  & $b_{000}$ & $b_{100}$ & $b_{010}$ & $b_{110}$ & $b_{001}$ & $b_{101}$ & $b_{011}$ & $b_{111}$ \\ \hline
\multirow{2}{*}{1.0}&\multirow{2}{*}{1.0} & predicted & 1.00 & 0.50 & 0.27 & 0.27 & 0.27 & 0.27 & 0.19 & 0.20 \\
&                                         &  measured & 1.00 & 0.50 & 0.25 & 0.24 & 0.25 & 0.24 & 0.15 & 0.16 \\\hline
\multirow{2}{*}{1.0}&\multirow{2}{*}{0.5} & predicted & 1.00 & 0.61 & 0.35 & 0.34 & 0.17 & 0.18 & 0.15 & 0.16 \\
&                                         &  measured & 1.00 & 0.62 & 0.33 & 0.34 & 0.17 & 0.18 & 0.13 & 0.15 \\\hline
\multirow{2}{*}{0.5}&\multirow{2}{*}{0.5} & predicted & 1.00 & 0.74 & 0.21 & 0.22 & 0.21 & 0.22 & 0.15 & 0.15 \\
&                                         &  measured & 1.00 & 0.75 & 0.23 & 0.25 & 0.23 & 0.25 & 0.14 & 0.15 \\\hline
\multirow{2}{*}{0.5}&\multirow{2}{*}{0.25}& predicted & 1.00 & 0.87 & 0.27 & 0.28 & 0.13 & 0.13 & 0.12 & 0.12 \\
&                                         &  measured & 1.00 & 0.83 & 0.31 & 0.34 & 0.16 & 0.17 & 0.13 & 0.13 \\
\hline\hline
\end{tabular}
\caption{Prediction of \eqref{alpha} versus the measurement obtained from the numerical experiments. 
In all cases, disturbance is introduced in the direction 1.
$b_{ijk}$ denotes the velocity of the computational cell located at $[ia^{(1)},ja^{(2)},ka^{(3)}]$ relative to the perturbed computational cell that is located at [0 0 0] (Figure \ref{fig:Kp-schematic}).}
\label{table:far_field}
\end{table}

Having defined $\alpha_{jk}^{(i)}$, $\beta_j^{(i)}$, and $\gamma_j^{(i)}$, $K\p^{(i)}$ can now be computed.
The two-way coupling force $F^{(i)}$ exerted at an arbitrary point leads to forces distributed at surrounding computational cells $F^{(i)}_j$ according to Eq. \eqref{force_dist}.
The induced velocity at the computational cell $j$ is then a result of the force exerted to the cell $j$ as well as those applied to surrounding cells $k$.
In other words, the velocity measured at a given computational cell $j$ can be thought as the superposition of several velocities, each created as a response to the force that is applied to the adjacent computational cell $k$, that can be quantified from Eq. \eqref{alpha}.
Simplifying Eq. \eqref{uc_ge} and \eqref{Kt} for a stationary ($C_{\rm t}^{(i)} = 1$) small force ($C_{\rm r} = 1$) at the steady state limit, the resulting velocity is
\begin{equation}
u_{{\rm c},j}^{(i)}  = -\frac{1}{ 3\pi\mu d\c K\c^{(i)}} \alpha^{(i)}_{jk} \beta_k^{(i)} F^{(i)}.
\label{vel_surrounding}
\end{equation}
Equation \eqref{vel_surrounding} neglects the velocity induced by $F^{(i)}$ in directions other than $i$. 
These perpendicular velocities are relatively small and are neglected in our analysis for the sake of simplicity. 
The disturbed velocities have an effect on the velocity at the location of particle according to Eq. \eqref{vel_int}, which in combination with Eq. \eqref{vel_surrounding} yields 
\begin{equation}
u\c^{(i)}  = -\frac{1}{ 3\pi\mu d\c K\c^{(i)}} \gamma_j^{(i)} \alpha^{(i)}_{jk} \beta_k^{(i)} F^{(i)}.
\label{vel_surrounding_2}
\end{equation}
Therefore, by taking
\begin{equation}
K\p^{(i)} = \gamma_j^{(i)}\alpha_{jk}^{(i)}\beta_k^{(i)},
\label{kp}
\end{equation}
we can account for the effect of the interpolation scheme on $K_{\rm t}^{(i)}$.
For a particle positioned exactly on top of a grid point, i.e., when all interpolation coefficients are zero except one, interpolation scheme does not affect $\bl u\c$ and $K\p^{(i)} = 1$.
This condition can be readily verified from Eq. \eqref{kp}, since $\alpha^{(i)}_{jj} = 1$, $\beta^{(i)}_1 = \gamma^{(i)}_1 = 1$, and $\beta^{(i)}_j = \gamma^{(i)}_j = 0$ for $j\ne 1$. 
Equation \eqref{kp} provides a generic relationship for $K\p^{(i)}$, which is valid at the limit of small cell Reynolds number for arbitrary cell geometry and interpolation scheme.

\subsection{Correction for the finite-Reynolds number}
To account for the change in the Stokes drag at Reynolds numbers larger than zero, we adopt the empirical relationship provided in \cite{clift2005bubbles} that is 
\begin{equation}
C_{\rm r} = 1 + 0.15{\rm Re}\c^{0.687},
\label{cr}
\end{equation}
where 
\begin{equation}
{\rm Re}\c = \frac{\|\bl u\c\| d\c}{\nu}
\label{rec}
\end{equation}
is the computational cell Reynolds number defined based the cell's velocity and its nominal diameter. 
Note that Eq. \eqref{cr} captures the effect of finite Reynolds number to a limited extent.
Although it partly accounts for the change in the Stokes drag, it does not account for the asymmetry in the disturbance field around a particle that invalidates Eq. \eqref{alpha} and the estimation of $K\p^{(i)}$.
Future refinement of Eq. \eqref{cr} may entail a more elaborate procedure that accounts for those effects by replacing or complementing the present estimation of $C_{\rm r}$. 

\subsection{Correction for the finite exposure}
A particle that is traveling through the grid spends a finite period within a computational cell. 
Thus, the period at which a computational cell is subjected to $F^{(i)}$ is limited by the exposure time of the particle to that computational cell. 
This limited exposure of the computational cell to $F^{(i)}$ is not captured by the time-dependent term in Eq. \eqref{uc_ge} and must be accounted for separately. 
To show the need for this additional correction, consider a particle with high velocity that spends an infinitesimal time in each computational cell. 
Since that infinitesimal time is not sufficient to accelerate the fluid within the computational cell, $\bl u\c \to 0$, obviating the need for any correction. 
However, Eq. \eqref{uc_ge} that is formulated in the Lagrangian frame of the particle is integrated concurrent with the particle (Eq. \eqref{particle_eq}).
Thus, the recorded time is not reset with the entrance of a particle to a computational cell but monotonically increases, yielding a finite $\bl u\c$.
To account for the finite time at which particle passes through a computational cell, we employ $C_{\rm t}^{(i)}$ to perform an additional correction. 

Solving Eq. \eqref{uc_ge} for a small force on top of the computational cell yields
\begin{equation}
u\c^{(i)}(t)= \left(1 - \exp{\left(-\frac{t-t_0}{\tau\c^{(i)}}\right)} \right) U\c^{(i)},
\label{transient_sol}
\end{equation}
where 
\begin{equation}
\tau\c^{(i)} = \frac{d\c^2}{12 \nu K\c^{(i)}}
\label{tauc}
\end{equation}
is the relaxation-time of the computational cell, $t_0$ is the time at which particle enters a computational cell, and $U\c^{(i)}=\lim_{t\to \infty}  u\c^{(i)}(t)$ is the terminal velocity of the computational cell.
According to Eq. \eqref{transient_sol}, the degree to which $\bl u\c$ is modified is time-dependent and its estimation requires one to keep track of $t_0$.
To eliminate this dependence and simplify the final result, we compute $C_{\rm t}^{(i)}$ such that the modification implied by Eq. \eqref{transient_sol} is accounted for in a time-averaged sense. 
By averaging Eq. \eqref{transient_sol} from $t_0$ to $t_0+\Delta t^{(i)}$, we obtain
\begin{equation}
C_{\rm t}^{(i)} = 1 - \frac{\tau\c^{(i)}}{\Delta t^{(i)}}\left(1 - \exp{\left(-\frac{\Delta t^{(i)}}{\tau\c^{(i)}}\right)}\right),
\label{ct}
\end{equation}
where
\begin{equation}
\Delta t^{(i)} = \frac{a^{(i)}}{|u\p^{(i)}|}
\label{ti}
\end{equation}
is the typical time for a particle to pass through a computational cell.
Asymptotically, Eq. \eqref{ct} implies that as the particle travels faster and $\Delta t^{(i)} \to 0$, $C_{\rm t}^{(i)} \to \Delta t^{(i)}/(2\tau\c^{(i)})$ and since $u^{(i)}\c \propto C_{\rm t}^{(i)}$, $\bl u\c \to 0$.
That is a limit that renders the need of having a correction scheme in the first place.
On the other hand, for very slowly moving particles $\Delta t^{(i)} \to \infty$ and $C_{\rm t}^{(i)} \to 1$. 
This is the limit that was exploited in the previous sections when we derived relationships for $K\c^{(i)}$, $K\p^{(i)}$, and $C_{\rm r}$.
In the next section, we show how all these steps can be combined to obtain a correction scheme. 

\subsection{The complete algorithm} \label{sec:comp}
The implementation of the present correction depends on the underlying time integration scheme.
Here, we assume an explicit time integration scheme, where $\bl u\p$ and $\bl u\c$ are given from the previous time step, and provide a procedure for computing $\dot u\p^{(i)}$ and $\dot u\c^{(i)}$ at every time step.
The following algorithm provides a general overview of how that is accomplished in our implementation:
\begin{enumerate}
\item Compute the disturbed fluid velocity at the location of particle $u_{\rm d}^{(i)}$. 
\item Compute the corrected fluid velocity $u\f^{(i)}$ using Eq. \eqref{ufe_def}.
\item Compute forces exerted to the particle $F^{(i)}$ using Eq. \eqref{F_exp}.
\item Compute $\dot u\p^{(i)}$ using Eq. \eqref{particle_eq}.
\item Based on the grid size $[a^{(1)}, a^{(2)}, a^{(3)}]$, compute $d\c$, $d_{\rm s}$, and $d_{\rm n}^{(i)}$ (Eqs. \eqref{dc} and \eqref{ds_dn}) and thereby $K\c^{(i)}$ using Eq. \eqref{kc}.
Note for a homogeneous grid $K\c^{(i)}$ is simply $0.516$.
\item From the location of the particle relative to the surrounding grid points (Figure \ref{fig:Kp-schematic}), compute $r_{jk}$ and $\theta_{jk}^{(i)}$ and thereby $\alpha_{jk}^{(i)}$ using Eq. \eqref{alpha}.
\item Based on the interpolation coefficients $\beta_j^{(i)}$ and $\gamma_j^{(i)}$ and $\alpha_{jk}^{(i)}$, compute $K\p^{(i)}$ using Eq. \eqref{kp}.
\item Compute $\rm Re\c$ and subsequently $C_{\rm r}$ using Eqs. \eqref{rec} and \eqref{cr}, respectively. 
\item Use Eqs. \eqref{tauc} and \eqref{ti} to compute $\tau\c^{(i)}$ and $\Delta t^{(i)}$, respectively, and substitute them in Eq. \eqref{ct} to obtain $C_{\rm t}^{(i)}$.
\item From $K\c^{(i)}$, $K\p^{(i)}$, $C_{\rm r}$, and $C_{\rm t}^{(i)}$, compute $K_{\rm t}^{(i)}$ using Eq. \eqref{Kt}.
\item Compute $\dot u\c^{(i)}$ using Eq. \eqref{uc_ge} (the history can be added as outlined in \ref{app:history}). 
\end{enumerate}

In practice, the entire correction scheme is broken into three independent subproblems in three directions, each time solving for a direction $i$. 
In other words, all the equations governing $\dot u\c^{(i)}$ are written solely in terms of $u\p^{(i)}$ and $u\c^{(i)}$ (the only exception is Eq. \eqref{rec} where $\|\bl u\c\|$ appears). 
Thus, corrections are essentially decoupled in low Reynolds number limit.

The order at which $\dot u\c^{(i)}$ and $\dot u\p^{(i)}$ are computed can be altered. 
The difference of such alteration, however, is of the same order as the error of the time integration scheme.
The initial condition for the computational cell velocity is zero, i.e. $\bl u\c(0) = 0$, since the fluid is not disturbed by the particle at $t=0$.
A simplified version of this correction scheme is described in \ref{sec:simplified}, where $\bl u\c$ is governed by an algebraic rather than a differential equation. 
In what follows, we evaluate the accuracy of this correction scheme under various conditions. 

\section{Results} \label{sec:result}
To evaluate the accuracy of our scheme under various conditions, we adopt four sets of test cases.
The first set of test cases is performed on an isotropic grid, where we examine the effect of the grid size and the Reynolds and Stokes numbers on the velocity of a settling particle.
The second set is similar to the first set with the difference that it is performed on anisotropic grids.
The third set involves the settling of two particles, where we examine the accuracy of our scheme in predicting the settling of the pairs of spheres.
The fourth and final set is a decaying homogeneous isotropic turbulence flow laden with particles, where we compare our scheme against particle-resolved simulations.

Our computational framework contains two components where the fluid and particle equations are solved in a Eulerian and a Lagrangian frame, respectively.
To simulate the fluid flow, we solve the incompressible Navier-Stokes equations by resolving all scales of the flow in the absence of particles.
For particle motion, we perform Lagrangian tracking of particles (Eq. \eqref{particle_eq}) and treat them as points without satisfying the no-slip condition on the surface of the particles.
Fluid equations are discretized using a finite-volume formulation and implemented in an in-house code that is staggered and second order accurate in space.
Second order Runge-Kutta scheme is used for the time integration along with a projection method to satisfy conservation of mass at each iteration of the Runge-Kutta scheme \cite{chorin1967numerical, kim1985application}. 
The position and velocity of the Lagrangian particles are integrated using the same time integration scheme as the fluid. 
The fluid and particle equations are fully coupled to ensure 2\textsuperscript{nd} order accuracy in time.
Unless stated otherwise, only the Stokes drag and buoyancy force are considered to model $\bl F$ in Eq. \eqref{F_exp}.
The distribution of the coupling force $\bl F$ to the fluid and also the computation of the disturbed fluid velocity at the location of the particle $\bl u_{\rm d}(\bl x\p)$ is performed using a trilinear interpolation scheme \cite{franklin2010high}.
Thus, both interpolation coefficients $\beta_j^{(i)}$ and $\gamma_j^{(i)}$ in Eqs. \eqref{force_dist} and \eqref{vel_int}, respectively, are identical in our framework. 
All simulations are performed using an in-house code \cite{esmaily2015scalable}.
This code has been employed for characterization of particle clustering \cite{esmaily2016CPT, esmaily2017modal} and validated against experiments \cite{esmaily2017computational}.

In the first three case studies considered below, one or two particles settle under gravity, and their velocity is measured and compared against the analytical solution to establish the accuracy of the present correction scheme.
The shared parameters among all these cases are the number of grid points, the boundary conditions of the computational domain, the direction of gravity, and the duration of the simulation.
The computational domain is a periodic box with $128^3$ grid points.
The size of the computational domain at this number of grid points is large enough such that disturbances in the far-field are small and the boundary conditions have minimal influence on the results.
Following the recommendation in reference \cite{horwitz2016accurate}, the gravity vector is selected such that $g^{(2)}/g^{(1)} = (1+\sqrt 5)/2$ and  $g^{(3)}/g^{(1)} = \exp(1)$.
With this choice, particles sweep through various points relative to the computational cells, ensuring the correction scheme is examined for different interpolation coefficients.
To ensure reported results are not affected by the integration period, all simulations are continued for at least $40\tau\p$, where 
\begin{equation}
\tau\p = \frac{\rho\p d\p^2}{18\nu}
\label{taup}
\end{equation}
is the particle relaxation time.
All the errors are computed once the initial transient period is passed, which typically entails averaging over ensembles obtained from $t > 10 \tau\p$. 
In some cases, e.g., low Stokes number cases below, the integration is continued up to $800\tau\p$ to get fully converged ensemble-averaged quantities. 
 
The cases considered in the first three case studies, which involve particle settling, are fully defined based on three non-dimensional parameters.
The first parameter, which has three independent components, is the ratio of the particle diameter to the grid size, defined as 
\begin{equation}
\Lambda^{(i)} = \frac{d\p}{a^{(i)}}.
\label{lambda}
\end{equation}
The second parameter is the particle Reynolds number, which is defined based the reference settling velocity $\bl u_{\rm r}$ as
\begin{equation}
{\rm Re\p} = \frac{\|\bl u_{\rm r}\|d\p}{\nu}.
\label{Rep}
\end{equation}
Note that $\bl u_{\rm r}$ is related to the gravitational acceleration by $\bl u_{\rm r} = \left(1- \rho\p/\rho\f\right)^{-1} \tau\p \bl g$.
The last parameter is the particle Stokes number.
The Stokes number is typically defined as the ratio of particle relaxation to the dissipation time scale of the flow.
Provided that the dissipation time scale of the flow is of the order of the smallest grid size squared divided by the kinematic viscosity of the flow, we define
\begin{equation}
{\rm St} = \nu \tau\p\min\left(a^{(i)}\right)^{-2}.
\label{St}
\end{equation}
Based on these three non-dimensional parameters, one can show $\rho\p/\rho\f = 18\max\left(\Lambda^{(i)}\right)^{-2}{\rm St}$, which is always larger than one in all cases considered below. 

To evaluate the accuracy of the present scheme we rely on the following metrics. 
The velocity of the particle, obtained from the simulation, is decomposed to 
\begin{equation}
\bl u\p^\|(t) = \frac{\bl u\r \cdot \bl u\p(t)}{\|\bl u\r\|^2} \bl u\r,
\label{up_parallel}
\end{equation}
which is the velocity parallel to the reference velocity $\bl u_{\rm r}$, and
\begin{equation}
\bl u\p^\bot(t) = \bl u\p(t) - \bl u\p^\|(t),
\label{up_normal}
\end{equation}
which is the velocity normal to $\bl u_{\rm r}$.
Based on these two velocities, we measure 
\begin{equation}
e^\| = \frac{\overline{\bl u\p^\|(t) \cdot \bl u\r}}{\|\bl u\r\|^2} - 1,
\label{e_settling}
\end{equation}
which is the error in the settling velocity that can be either positive or negative, and 
\begin{equation}
e^\bot = \frac{\overline{\|\bl u\p^\bot(t)\|}}{\|\bl u\r\|},
\label{e_drift}
\end{equation}
which signifies the deviation of particle from a straight trajectory. 
In Eqs. \eqref{e_settling} and \eqref{e_drift}, $\overline{(\bullet)}$ denotes a time average. 
Additionally, to have a measure of overall error, we employ
\begin{equation}
e = \frac{\overline{\|\bl u\p(t) - \bl u\r\|}}{\|\bl u\r\|}.
\label{e_total}
\end{equation}

\subsection{The effect of grid size, Reynolds and Stokes numbers} \label{sec:settling}
The first set of test cases are similar to those discussed in \cite{horwitz2016accurate}.
These cases involve simulating a single particle settling under gravity in an isotropic grid with $\Lambda^{(1)}=\Lambda^{(2)}=\Lambda^{(3)}$. 
The grid size, the particle Reynolds number, and the Stokes number are varied according to Eqs. \eqref{lambda}, \eqref{Rep}, and \eqref{St}, respectively, and errors are computed according to Eqs. \eqref{e_settling}, \eqref{e_drift}, and \eqref{e_total}. 
In total 19 cases are simulated with results shown in Table \ref{table:uniform}.
To provide a reference, errors are also reported for the simulations run with no correction as well as with the correction proposed in \cite{horwitz2016accurate} using trilinear interpolation.

\begin{table}
\centering
\begin{tabular}{cccc|cc|cc|ccc}
  \hline \hline
  & & & & \multicolumn{2}{c|}{uncorrected} & \multicolumn{2}{c|}{H\&M} & \multicolumn{3}{c}{present correction}\\
 Case & $\Lambda^{(i)}$ & $\rm Re\p$ & $\rm St$ & $e^\|$ & $e^\bot$ & $e^\|$ & $e^\bot$ & $e^\parallel$ & $e^\bot$ & $e$ \\ \hline
U01&\g$ 0.01 $&$ 0.1  $&$ 10    $&$   0.17^\dagger $&$ 0.003  $&$  0.016 $&$ 0.002 $&$ -0.024 $&$ 0.054 $&$ 0.06 $\\
U02&\g$ 0.05 $&$ 0.1  $&$ 10    $&$   2.5  $&$ 0.006  $&$  0.16  $&$ 0.011 $&$ -0.13  $&$ 0.30  $&$ 0.32 $\\
U03&\g$ 0.1  $&$ 0.1  $&$ 10    $&$   6.2  $&$ 0.025  $&$  0.35  $&$ 0.010 $&$ -0.15  $&$ 0.36  $&$ 0.39 $\\
U04&\g$ 0.25 $&$ 0.1  $&$ 10    $&$  18    $&$ 0.14   $&$  1.0   $&$ 0.020 $&$  0.047 $&$ 0.38  $&$ 0.38 $\\
U05&\g$ 0.5  $&$ 0.1  $&$ 10    $&$  37    $&$ 0.42   $&$  2.4   $&$ 0.063 $&$  0.43  $&$ 0.39  $&$ 0.58 $\\
U06&\g$ 1    $&$ 0.1  $&$ 10    $&$  75    $&$ 1.0    $&$  5.1   $&$ 0.22  $&$  0.83  $&$ 0.44  $&$ 1.0  $\\
U07&\g$ 2    $&$ 0.1  $&$ 10    $&$ 150    $&$ 2.0    $& --$^\ddagger$ & -- &$  1.2   $&$ 0.92  $&$ 1.9  $\\
U08&\g$ 4    $&$ 0.1  $&$ 10    $&$ 290    $&$ 3.2    $& -- & -- &$  1.5   $&$ 2.8   $&$ 5    $\\
U09&\g$ 5    $&$ 0.1  $&$ 10    $&$ 360    $&$ 3.6    $& -- & -- &$  1.7   $&$ 5.2   $&$ 7.5  $\\\hline
U10&$ 1    $&\g$ 0.05 $&$ 10    $&$  76    $&$ 1.9    $&$  5.5   $&$ 0.37  $&$ -0.31  $&$ 0.40  $&$ 0.78 $\\
U11&$ 1    $&\g$ 0.25 $&$ 10    $&$  71    $&$ 0.45   $&$  3.4   $&$ 0.13  $&$  2.7   $&$ 0.98  $&$ 2.9  $\\
U12&$ 1    $&\g$ 0.5  $&$ 10    $&$  65    $&$ 0.25   $&$  0.65  $&$ 0.093 $&$  4.3   $&$ 2.0   $&$ 4.7  $\\
U13&$ 1    $&\g$ 1    $&$ 10    $&$  56    $&$ 0.19   $& -- & -- &$  6     $&$ 4.2   $&$ 7.3  $\\
U14&$ 1    $&\g$ 5    $&$ 10    $&$  27    $&$ 0.27   $& -- & -- &$  4.6   $&$ 8.1   $&$ 9.4  $\\
U15&$ 1    $&\g$ 10   $&$ 10    $&$  16    $&$ 0.28   $& -- & -- &$  2.3   $&$ 5.7   $&$ 6.2  $\\\hline
U16&$ 1    $&$ 0.1  $&\g$  0.25 $&$  76    $&$ 8.2    $&$ -4.7   $&$ 0.86  $&$  0.43  $&$ 0.86  $&$ 1.4  $\\
U17&$ 1    $&$ 0.1  $&\g$  0.5  $&$  76^\dagger $&$ 7.3    $&$ -1.7   $&$ 0.89  $&$  0.42  $&$ 0.84  $&$ 1.4  $\\
U18&$ 1    $&$ 0.1  $&\g$  1    $&$  76^\dagger $&$ 5.9    $&$  1.4   $&$ 0.61  $&$  0.41  $&$ 0.80  $&$ 1.3  $\\
U19&$ 1    $&$ 0.1  $&\g$  5    $&$  75    $&$ 2.0    $&$  4.5   $&$ 0.39  $&$  0.22  $&$ 0.54  $&$ 0.97 $\\
   \hline\hline
\end{tabular}
\caption{The error in the simulated velocity of a single particle settling under gravity on an isotropic grid. 
The effect of the grid size $\Lambda^{(i)}$, particle Reynolds number $\rm Re\p$, and the Stokes number $St$ are shown on the error in the settling velocity $e^\|$, error in the drift velocity $e^\bot$, and the overall error $e$.
The present correction scheme is compared with the case with no correction and H\&M, which is the correction proposed in \cite{horwitz2016accurate}.
$e$ is not reported for uncorrected cases since it is almost identical to $e^\|$.
$^\dagger$ The error reported for some of the uncorrected cases is slightly different from those reported in \cite{horwitz2016accurate} since the time averaging is performed for a longer period in this study. 
$^\ddagger$ Results were not available for these cases.}
\label{table:uniform}
\end{table}

There are a number of trends that are evident from Table \ref{table:uniform}.
If we consider the uncorrected cases, the error is mainly a function of $\Lambda^{(i)}$, increasing as $\Lambda^{(i)}$ increases.
The error is dominated by the wrong estimation of the settling velocity as $e^\| \gg e^\bot$. 
The present correction method is the best at reducing this part of the error, providing a much better estimation of the settling velocity. 
In fact, $e^\|$ after applying the correction is reduced to the same order of magnitude as $e^\bot$. 
This reduction in $e^\|$ has led to an overall error that is less than 2\% for most cases. 
The exceptions are U08 and U09, which have extremely large $\Lambda^{(i)}$ with uncorrected errors in the order of 300\%, and U11 to U15, which have relatively large $\rm Re\p$.
Considering cases U16 to U19 shows that the Stokes number has minimal effect on the errors associated with the uncorrected or the present correction scheme. 
In general, the present correction scheme reduces the error at low $\rm Re\p$ by approximately two orders of magnitude for $\Lambda^{(i)}>0.1$, which is when a correction is needed.

As cases U11 to U15 show, our correction method is more effective at reducing error at lower particle Reynolds numbers.
Based on Eqs. \eqref{stokes} and \eqref{uc_sim}, $\|\bl u\c\|d\c \propto \|\bl u\p\|d\p$, thus the Reynolds of the computational cell $\rm Re\c$ scales similar to $\rm Re\p$.
For most cases considered in this study $\rm Re\p = 0.1$, yielding a sufficiently low $\rm Re\c$ to assume a Stokes flow around the computational cell.
The significance of having a Stokes flow around the computational cell pertains to the validity of Eq. \eqref{alpha} in estimating $\alpha_{jk}^{(i)}$, as it directly affects $K\p^{(i)}$.
It is the breaking of the flow symmetry at higher Reynolds numbers that is not captured by Eq. \eqref{alpha}, explaining relatively larger errors observed in cases U11 to U15.
Considering the largest Reynolds number case U15 for instance, $e^\bot$ relative to the corresponding uncorrected case is the largest, suggesting a significant deviation of $\bl u\p$ from the direction of gravity.
Despite its lower performance, the present correction scheme reduces the error by more than 60\% at such high particle Reynolds number.
A further decrease in error would require an alternative formulation of Eq. \eqref{alpha} such that it would differentiate between grid points upstream and downstream of the particle by assigning larger weights to grid points in the wake of the particle.
Accounting for this anisotropy in the disturbed velocity field is essential for obtaining a more accurate correction scheme at $\rm Re\p \ge \mathcal O(1)$, warranting further refinement of the present approach in the future.

\begin{figure}
\begin{center}
\includegraphics[width=0.6\textwidth]{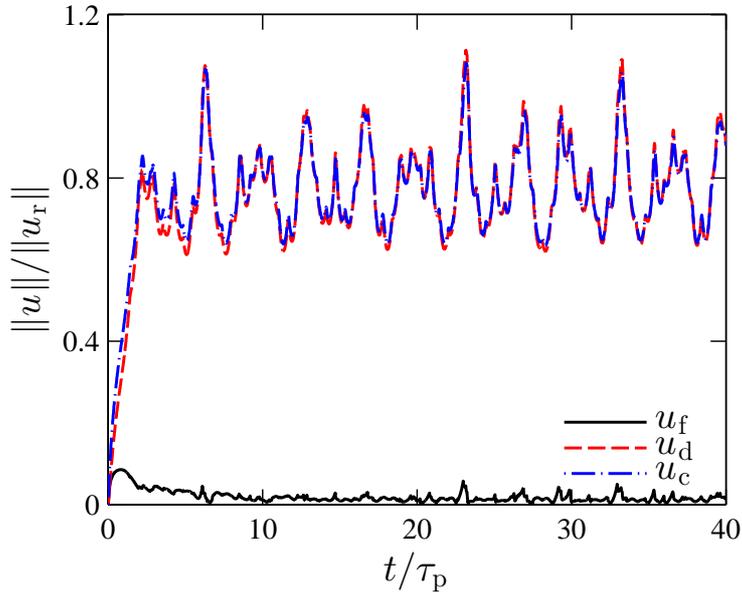}
\caption{The undisturbed fluid velocity for a settling particle estimated by the present correction scheme $\|\bl u\f\|$ (solid black) and the corresponding disturbed fluid velocity at the location of the particle $\|\bl u_{\rm d}\|$ (dashed red).
The velocity of the computational cell $\|\bl u\c\|$ (dash-dot blue) is also shown as a function of time. 
Results correspond to case U06 in Table \ref{table:uniform} (color online.)}
\label{fig:uc}
\end{center}
\end{figure}

To further analyze the present correction scheme, the disturbed, the undisturbed, and the computational cell velocity are shown in Figure \ref{fig:uc} for case U06 from Table \ref{table:uniform}.
Since the Stokes drag model is based on the undisturbed fluid velocity, the ideal correction scheme should produce $\bl u\f = 0$.
The interpolated velocity at the location of particle $\bl u_{\rm d}$ is far from zero since the fluid is dragged by the particle owing to the two-way coupling force.
However, once the velocity of the computational cell $\bl u\c$, obtained from the present correction scheme, is subtracted from $\bl u_{\rm d}$, a new estimate of fluid velocity is obtained, i.e., $\bl u\f$, that is much closer to zero.
The significant fluctuations in $\bl u_{\rm d}$, which is a result of the particle traversing through the computational cells, is well captured by $\bl u\c$ via its dependence on $K\p^{(i)}$.

\begin{figure}
\begin{center}
\includegraphics[width=0.6\textwidth]{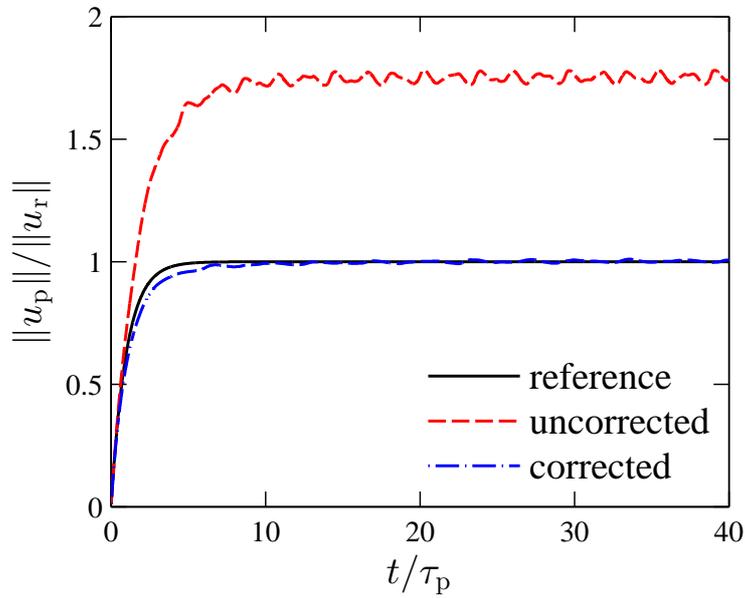}
\caption{The velocity of a settling particle as function of time, predicted analytically (solid black) and computed in a two-way coupled simulation without correction (dashed red) and with the present correction (dash-dot blue).
These results are based on case U06 in Table \ref{table:uniform} (color online.)}
\label{fig:up}
\end{center}
\end{figure}

The accurate prediction of the particle velocity $\bl u\p$ relies on the correct estimation of $\bl u\f$ (Figure \ref{fig:up}).
A two-way coupled simulation with no correction, which takes $\bl u_{\rm d}$ as an estimate of $\bl u\f$, produces a large error in estimation of $\bl u\p$ since $\bl u_{\rm d}$ is significantly different from the undisturbed velocity $\bl u\f$, which is zero in this case. 
The predicted $\bl u\p$ that is shown in Figure \ref{fig:up} is 75\% larger than the reference velocity (case U06 in Table \ref{table:uniform}). 
On the other hand, the present scheme corrects $\bl u\f$ (Figure \ref{fig:uc}), thus providing a much better approximation of $\bl u\p$ (Figure \ref{fig:up}).
In the transient period, there is a small error in its prediction, which is largely caused by neglecting the history effects. 
Following the procedure in \ref{app:history} to include the history term will reduce the error in the transient period. 
Once the transient period is passed, however, the corrected scheme predicts a $\bl u\p$ that is less than 1\% different from the reference velocity.
 
\subsection{The effect of grid anisotropy}
To demonstrate the validity of the present correction scheme on anisotropic grids, we repeat simulations with a settling particle from Section \ref{sec:settling} but using non-identical $\Lambda^{(i)}$'s. 
The particle Reynolds and Stokes numbers are kept constant at 0.1 and 10 in these simulations, reproducing conditions similar to U06. 
The grid aspect ratio, defined as the ratio of the largest to the smallest $\Lambda^{(i)}$, is varied between 2 to 20 to represent a variety of grid spacing that may be encountered in a channel flow simulation laden with particles.
The higher aspect ratio is typical of grid cells near walls, whereas the lower aspect ratio is representative of grid spacing in the center of the channel. 
In total, 10 cases are simulated and the results are shown in Table \ref{table:nonuniform}.

\begin{table}
\centering
\begin{tabular}{ccccc|ccc|ccc}
  \hline \hline
 & & & & & \multicolumn{3}{c|}{uncorrected} & \multicolumn{3}{c}{corrected}\\
Cell shape & Case & $\Lambda^{(1)}$ & $\Lambda^{(2)}$ & $\Lambda^{(3)}$ & $e^\|$ & $e^\bot$ & $e$ & $e^\parallel$ & $e^\bot$ & $e$ \\ 
\hline \multirow{3}{*}{\includegraphics[width=0.08\textwidth]{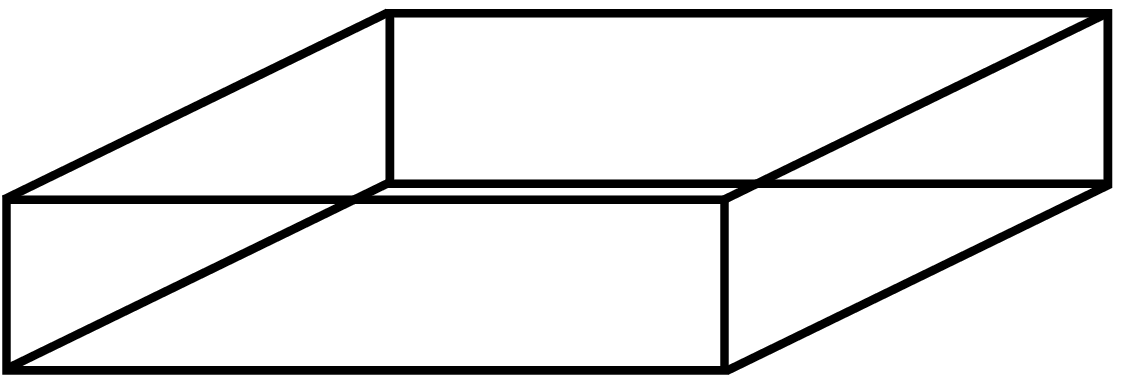}} 
&N01&$ 2   $&$ 1   $&$ 1    $&$ 92  $&$  5.5 $&$ 93  $&$  0.6  $&$ 0.69 $&$ 1.4 $\\
&N02&$ 2   $&$ 0.5 $&$ 0.5  $&$ 52  $&$  5.9 $&$ 53  $&$ -0.86 $&$ 0.73 $&$ 1.5 $\\
&N03&$ 5   $&$ 0.5 $&$ 0.5  $&$ 56  $&$  8.8 $&$ 57  $&$ -2.0  $&$ 1.8  $&$ 2.9 $\\
\hline \multirow{3}{*}{\includegraphics[width=0.08\textwidth]{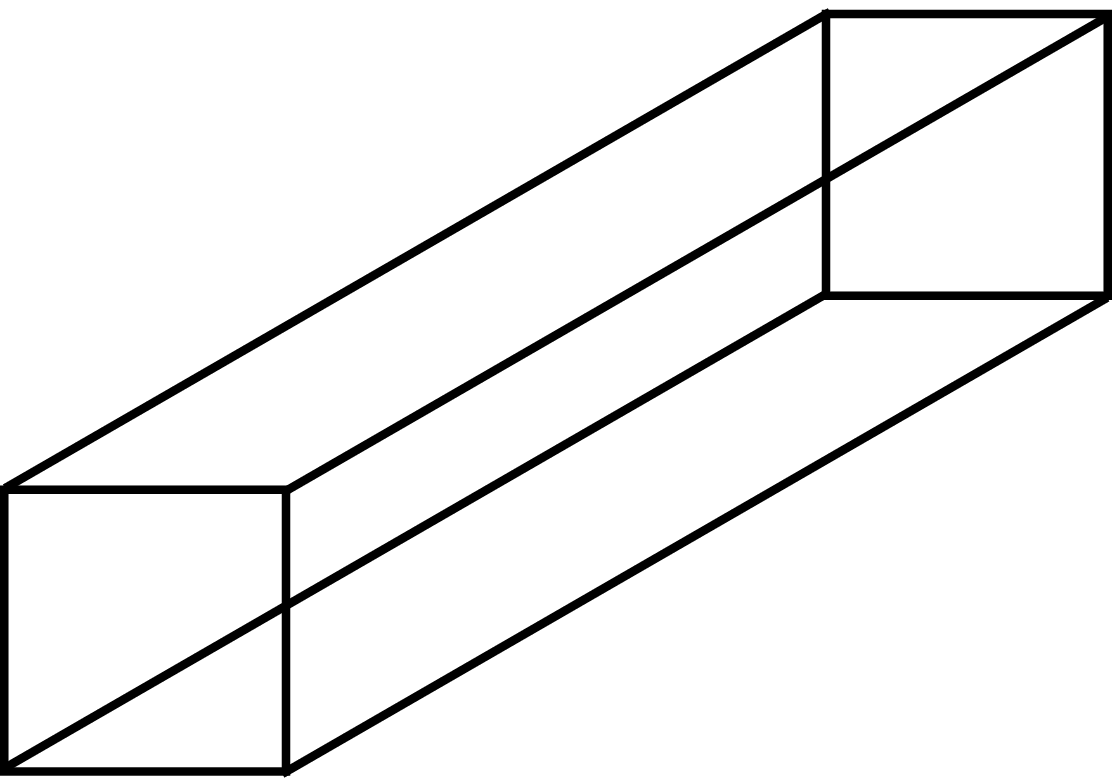}} 
&N04&$ 1   $&$ 1   $&$ 0.5  $&$ 59  $&$  4.8 $&$ 59  $&$  1.5  $&$ 0.38 $&$ 1.6 $\\
&N05&$ 2   $&$ 2   $&$ 0.5  $&$ 84  $&$ 11   $&$ 85  $&$  0.34 $&$ 0.81 $&$ 1.3 $\\
&N06&$ 2   $&$ 2   $&$ 0.2  $&$ 48  $&$  8.7 $&$ 49  $&$ -3.1  $&$ 3.3  $&$ 4.7 $\\
\hline \multirow{4}{*}{\includegraphics[width=0.08\textwidth]{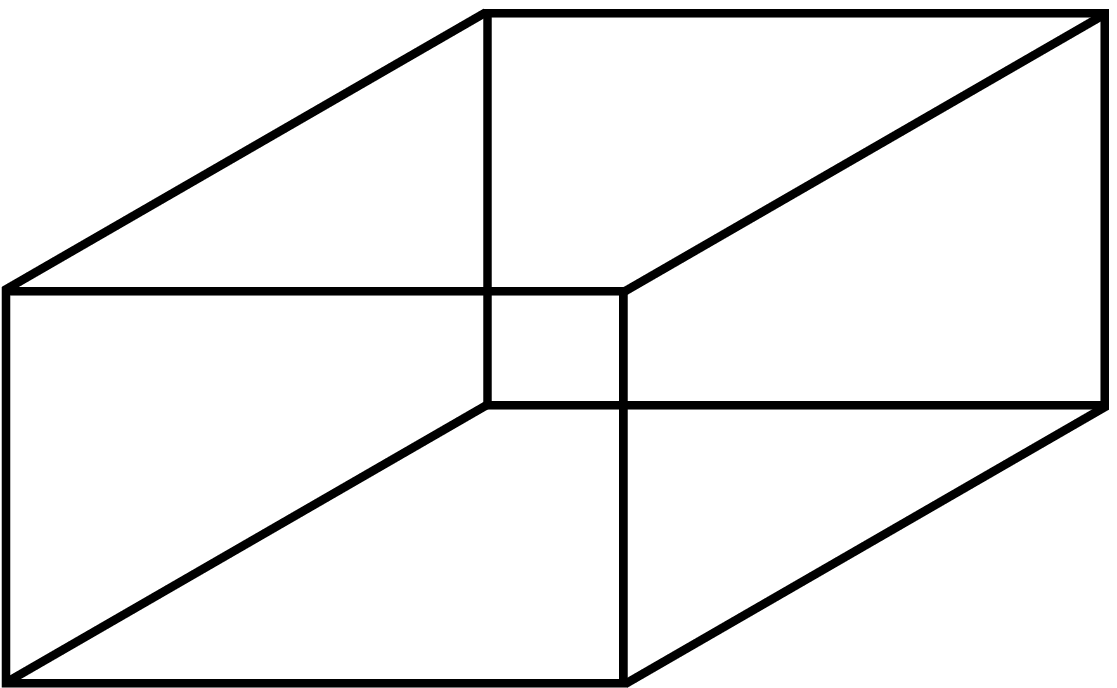}} 
&N07&$ 2   $&$ 1   $&$ 0.5  $&$ 69  $&$  7.7 $&$ 70  $&$  0.51 $&$ 0.63 $&$ 1.3 $\\
&N08&$ 2   $&$ 1   $&$ 0.25 $&$ 48  $&$  7.3 $&$ 48  $&$ -1.5  $&$ 1.1  $&$ 2.1 $\\
&N09&$ 4   $&$ 2   $&$ 0.2  $&$ 52  $&$  9.8 $&$ 53  $&$ -3.5  $&$ 6.0  $&$ 7.3 $\\
&N10&$ 4   $&$ 1   $&$ 0.25 $&$ 51  $&$  8.8 $&$ 52  $&$ -0.85 $&$ 2.3  $&$ 3.0 $\\
\hline\hline
\end{tabular}
\caption{The error in prediction of the velocity of a single particle settling under gravity on rectilinear anisotropic grids. 
Three classes of grids are examined with tabulated $\Lambda^{(i)}$.
The present correction scheme is compared against the case with no correction in terms of error in the settling velocity $e^\|$ (Eq. \eqref{e_settling}), error in drifting from a straight path $e^\bot$ (Eq. \eqref{e_drift}), and the total error $e$ (Eq. \eqref{e_total}).}
\label{table:nonuniform}
\end{table}

Consistent with the isotropic grid cases, these anisotropic cases show that the overall error before correction is primarily a function of the grid volume rather the smallest dimension of the grid.
In other words, the error increases proportional to $(\Lambda^{(1)}\Lambda^{(2)}\Lambda^{(3)})^{1/3} \propto d\p/d\c$ rather than $\max(\Lambda^{(i)})$. 
For example, cases U06 and N07 have identical $d\p/d\c$ with similar $e$ before correction, whereas cases U07 and N07 have identical $\max(\Lambda^{(i)})$ with significantly different $e$.
This observation is consistent with the formulation of present correction scheme.
Considering Eq. \eqref{K_def} with a constant $K_{\rm t}^{(i)}$ as the simplified form of the present correction, the required adjustment in the Stokes drag is proportional to $d\p/d\c$, suggesting that the error also grows proportional to $d\p/d\c$.

In comparison to the isotropic grid, simulation performed on anisotropic grid with no correction at similar $d\c$ produce larger $e^\bot$ (Tables \ref{table:uniform} and \ref{table:nonuniform}).
Comparing N01 and N03 that have roughly similar $d\c$ shows that the ratio of $e^\bot/e^\|$ increases with the grid aspect ratio.
After applying the correction scheme, both errors reduce significantly. 
Even at the largest aspect ratio, i.e., case N09 with an aspect ratio of 20, the overall error $e$ is reduced from 53\% to 7.3\%, showing 87\% reduction in error.
The reduction in error is more dramatic at lower aspect ratios, where the error is reduced from over roughly 50\% to less than 5\% when the present correction scheme is employed in these anisotropic grid cases. 

\subsection{Settling of two-particles}
In this section, we consider two particles that are falling side-by-side under gravity and compute their velocity and compare it against an analytical solution.
The analytical solution to this problem can be expressed as a modification to the settling velocity of a single particle under gravity \citep{Stimson-1926,Batchelor-1972,Batchelor_Green-1972}
The degree to which the settling velocity changes depends on the distance between the two particles, which is denoted here by $l$. 
If they are at an infinite distance, their settling velocity will correspond to a single particle, whereas, at a distance of order $d\p$ they fall faster as a result of the disturbance field introduced by the neighboring particle. 
The largest modification in the settling velocity occurs when $l=d\p$, i.e. when two particles are touching, at which $\|\bl u_{\rm r}\| = 1.38U_{\rm s}$ with $U_{\rm s}$ being the settling velocity of a single particle under identical conditions.

Under a free fall, two side-by-side particles experience not only a drag force along $-\bl u\p$, but also a repulsive force normal to $\bl u\p$ that separates two particles from each other. 
As a result, for the pair of particles that are not constrained, $l$ will change with time, changing their settling velocity. 
The time variation of $l$ is contrary to the underlying assumption of the analytical solution, which assumes particles to be at a fixed relative position. 
Therefore, to allow for a direct comparison, we constrain particles in our simulation to travel on two straight lines, which are parallel to each other with a distance $l$.
The resulting settling velocity is measured from these simulations and compared against the analytical solution in Table \ref{table:two_prt}.

\begin{table}
\centering
\begin{tabular}{cccccc|cc|cc}
\hline \hline
 & & & & &  &\multicolumn{2}{c|}{uncorrected} & \multicolumn{2}{c}{corrected} \\
Case & $\bl \Lambda$ & $\rm Re\p$ & $\rm St$ & $l/d\p$ & $\|\bl u_{\rm r}\|/U_{\rm s}$ & $\| \bl u\p \|/U_{\rm s}$ & $e^\|$ & $\|\bl u\p \|/U_{\rm s}$ & $e^\|$ \\ \hline
T01&\g$\bl 0.5 $&$ 0.1  $&$ 10    $& 1 & 1.38  & 1.667 &  21  & 1.313 &  $-4.9$\\
T02&\g$\bl 1   $&$ 0.1  $&$ 10    $& 1 & 1.38  & 2.139 &  55  & 1.413 &  $ 2.4$\\
T03&\g$\bl 2   $&$ 0.1  $&$ 10    $& 1 & 1.38  & 2.821 & 104  & 1.363 &  $-1.3$\\
T04&\g$\bl 4   $&$ 0.1  $&$ 10    $& 1 & 1.38  & 4.186 & 203  & 1.308 &  $-5.3$\\\hline
T05&\g$[2,1,1   ]$&$ 0.1  $&$ 10  $& 1 & 1.38  & 2.255 &  63  & 1.356 &  $-1.7$\\
T06&\g$[1,1,0.5 ]$&$ 0.1  $&$ 10  $& 1 & 1.38  & 1.931 &  40  & 1.369 &  $-0.8$\\
T07&\g$[2,1,0.5 ]$&$ 0.1  $&$ 10  $& 1 & 1.38  & 1.990 &  44  & 1.322 &  $-4.2$\\\hline
T08&$\bl 1   $&\g$ 0.05 $&$ 10    $& 1 & 1.38  & 2.162 &  57  & 1.407 &  $ 1.9$\\
T09&$\bl 1   $&\g$ 0.5  $&$ 10    $& 1 & 1.38  & 1.947 &  41  & 1.386 &  $ 0.4$\\\hline
T10&$\bl 1   $&$ 0.1  $&\g$  0.25 $& 1 & 1.38  & 2.139 &  55  & 1.406 &  $ 1.9$\\
T11&$\bl 1   $&$ 0.1  $&\g$  1    $& 1 & 1.38  & 2.135 &  55  & 1.405 &  $ 1.8$\\\hline 
T12&$\bl 1   $&$ 0.1  $&$ 10    $&\g 2 & 1.195 & 1.910 &  60  & 1.181 &  $-1.2$\\
T13&$\bl 1   $&$ 0.1  $&$ 10    $&\g 4 & 1.095 & 1.812 &  65  & 1.079 &  $-1.5$\\
\hline\hline
\end{tabular}
\caption{The settling velocity of two particles falling side-by-side under gravity obtained from the analytical solution $\|\bl u_{\rm r}\|$ \citep{Batchelor-1972}, two-way coupled simulation without correction and with the present correction scheme $\|\bl u\p\|$.
$U_{\rm s}$ is the magnitude of the velocity of a single particle settling under similar conditions.
$l$ is the distance between the centers of two particles.
Since particles are constrained to travel along a straight line, $e^\bot = 0$ and $e=|e^\||$ by construction, thus only $e^\|$ is reported.
Other parameters are defined in Table \ref{table:uniform}.}
\label{table:two_prt}
\end{table}

Analogous to the single particle settling cases, the predicted velocity of the pair of nearby particles is more accurate with the present correction, reducing errors that are typically more than 50\% to less than 5\% for cases considered in Table \ref{table:two_prt}. 
The strong dependence of the error on $\bl \Lambda$ and its relative independence from $\rm St$ and $\rm Re\p$ is also observed here. 
These cases demonstrate the accuracy of the present method in capturing the disturbance field on anisotropic grids (cases T05, T06, and T07), in the intermediate separation (cases T12 and T13), and for particles that are larger than the grid (cases T03 and T04). 

\begin{figure}
\begin{center}
\includegraphics[width=1.0\textwidth]{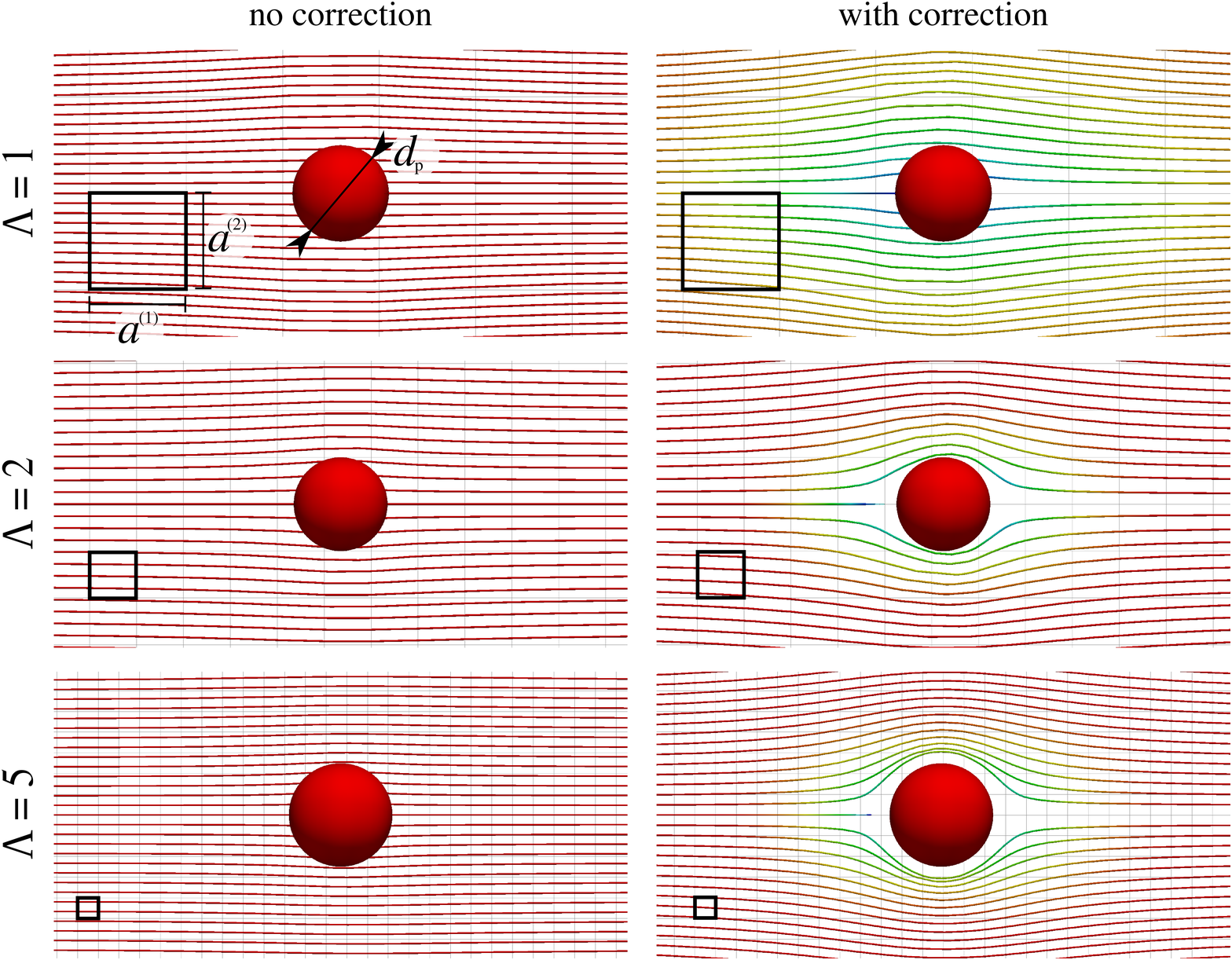}
\caption{The disturbance field created in the point-particle simulations. 
The streamlines are computed based on $\bl u_{\rm d} - \bl u\p$ for the uncorrected scheme (left) and corrected scheme (right).
The effect of various grid sizes relative to the particle diameter is shown from top to bottom corresponding to cases U06, U07, and U09 in Table \ref{table:uniform}, respectively (except $\bl g$ pointing in the x-direction in these simulations).
Particles are shown with a finite size for visualization purposes only as they are treated as points in these computations.}
\label{fig:streamlines}
\end{center}
\end{figure}

The correct estimation of the disturbance near particles requires a grid that has a similar size to that of the particle. 
The need for having a grid that is sufficiently small for capturing particle-particle interaction is demonstrated by the increase in the error from case T02 to T01.
As $\bl \Lambda \to 0$, the disturbance created by the particle due to the two-way coupling forces declines, leaving no mechanism for particles to influence each others motion (Figure \ref{fig:streamlines}).
In other words, a computational cell that is much larger than the particle can hardly differentiate a pair of particles that are at different distances from each other, as long as they both fit inside a computational cell.
Consequently, the provided estimate of $\|\bl u\p\|$ tends toward a case in which $l\approx d\c$, decreasing as $\bl \Lambda$ decreases.
For case T01, this error leads to a lower $\|\bl u\p\|$ that happens to cancel the over-estimation of $\|\bl u\p\|$ for the uncorrected scheme, producing a smaller overall error. 
The results collectively show that the present correction is effective at capturing the interaction of nearby particles at various physical and numerical conditions, as long as $\bl \Lambda$ is order one.

\subsection{Decaying homogeneous isotropic turbulence}
For the last case study, we consider decaying homogeneous isotropic turbulence laden with inertial particles. 
This test case is selected to evaluate the present correction scheme under more realistic conditions at which the flow is spatially varying, and particles follow complex trajectories. 
The parameters of the simulation are selected to match conditions with the particle-resolved simulation reported in \citep{subramaniam_et_al_2014}.
While simulations at $\rm Re_\lambda = 12$ and 27 are performed in that study, where $\rm Re_\lambda$ is the Reynolds number based on the Taylor microscale, only the lower Reynolds number results are shown in \citep{subramaniam_et_al_2014}. 
The results shown here, which correspond to $\rm Re_\lambda = 27$, are obtained through personal communication.
For the sake of completeness, we shortly describe these conditions and then compare the results of our point-particle simulations with that of particle-resolved simulations from \cite{subramaniam_et_al_2014}.

An isotropic turbulent flow is generated based on an energy spectrum that matches the initial spectrum in particle-resolved simulation, which is originally obtained from Pope's model \citep{pope-2000}, using Rogallo's procedure described in \citep{Rogallo-1981}.  
The diameter and density of particle are selected such that $d\p/\eta = 1$ and $\rm St_\eta = 1$, where $\eta$ is the Kolmogorov length scale and $\rm St_\eta$ is the particle Stokes number based on the Kolmogorov time scale.
The domain is seeded with particles that are initially at rest.
The number of particles is selected to obtain a volume fraction of $10^{-3}$.
The corresponding particle-resolved simulation has been performed on a $1152^3$ grid, yielding 12 grid points across a particle \cite{subramaniam_et_al_2014}.
In our point-particle simulations, on the other hand, the computational domain is discretized on a $96^3$ isotropic grid with $\bl \Lambda = 1$.
These choices of non-dimensional parameters lead to $\rho\p/\rho_{\rm f} = 18$, the mass fraction of 1.8\%, and $N\p = 1690$ as the number of particles in the entire domain.
Since the particle Reynolds number is relatively large in these calculations, $\bl F$ is modeled by the Stokes drag (Eq. \eqref{stokes}) that is corrected for the finite Reynolds number using an equation identical to Eq. \ref{cr} with $\rm Re\c$ replaced by $\rm Re\p$.

\begin{figure}
\begin{center}
\includegraphics[width=0.6\textwidth]{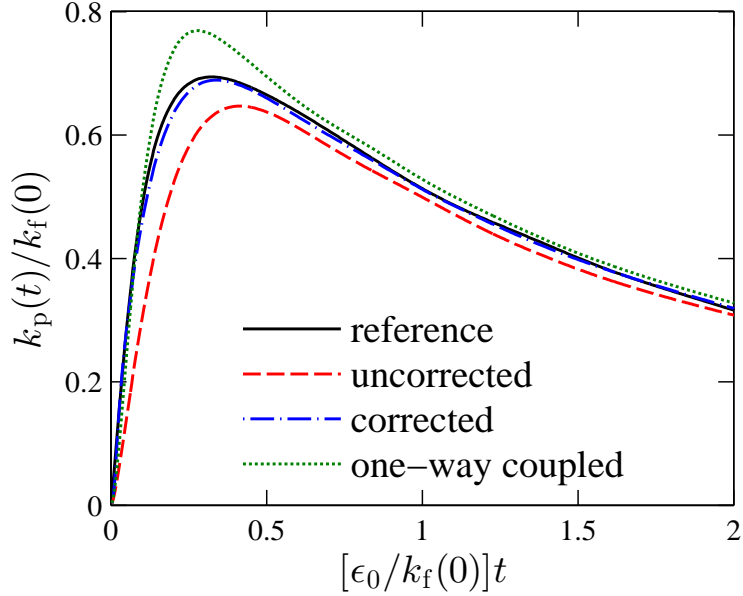}
\caption{The kinetic energy of particles in a decaying homogeneous isotropic turbulence as a function of time, from the particle-resolved simulation (solid black) \cite{subramaniam_et_al_2014}, the two-way coupled point-particle simulation with no correction (dashed red), the two-way coupled point-particle simulation with the present correction (dash-dot blue), and the one-way coupled point-particle simulation (dotted green).}
\label{fig:Kp}
\end{center}
\end{figure}

In total, three simulations were performed where the two-way coupling forces are neglected, the two-way coupling forces are included but no correction scheme is adopted, and the two-way coupling forces are added using the present correction scheme (Figure \ref{fig:Kp}). 
In these simulations, particles are initially at rest and
\begin{equation}
k\p(t) = \frac{1}{2 N\p}\sum_{i=1}^{N\p} \|\bl u_{{\rm p},i}(t)\|^2,
\label{ke_p}
\end{equation}
which is the particle kinetic energy, is zero at $t=0$. 
In Eq. \eqref{ke_p}, $\bl u_{{\rm p},i}(t)$ denotes the velocity of $i^{\rm th}$ particle. 
As time passes, particles experience a drag from the fluid and accelerate. 
As a result, early in this transient process, the fluid kinetic energy, defined as
\begin{equation}
k\f(t) = \frac{1}{V}\int_V\frac{1}{2} \|\bl u(\bl x,t)\|^2 {\rm d}V,
\label{ke_f}
\end{equation}
is transfered to the particles, increasing $k\p(t)$. 
In Eq. \eqref{ke_f}, $\bl u(\bl x,t)$ is the Eulerian representation of the fluid velocity and the integral is carried out over the entire computational domain that has volume $V$.
Note that $\bl u(\bl x\p,t) = \bl u_{\rm d}(t)$.
Concurrently, the fluid kinetic energy dissipates with rate $\epsilon(t)$ by the viscous forces, reducing the overall energy of the system.
Due to this dissipation, $k\f$ drops at the longer time such that the direction of energy transfer reverses and particles begin to transfer their energy to the fluid. 
This reversal of energy transfer direction coincides with $t$ at which $k\p$ becomes maximum (Figure \ref{fig:Kp}).
From the three point-particle simulations, only the simulation with the present correction accurately captured the time and magnitude associated with the maximum of $k\p(t)$.
The one-way coupled simulation is not energy conservative and neglects to subtract the energy that is transferred to the particles from the fluid.
Not removing energy from the system at the correct rate leads to a higher $k\f$ and consequently a higher peak for $k\p$ that is slightly time-advanced.
The uncorrected simulation, on the other hand, underestimates the drag on the particles, leading to a lower rate of energy transfer between fluid and particles in the acceleration phase, thereby predicting a lower peak for $k\p$ that is time-delayed.
With the addition of the present correction to the two-way coupled point-particle simulation, the drag force is computed with a good approximation.
As a result of this correct prediction, the rate of energy transfer is calculated correctly, leading to an accurate prediction of the magnitude of the peak of $k\p$ and the time at which it occurs.

\begin{figure}
\begin{center}
\includegraphics[width=0.6\textwidth]{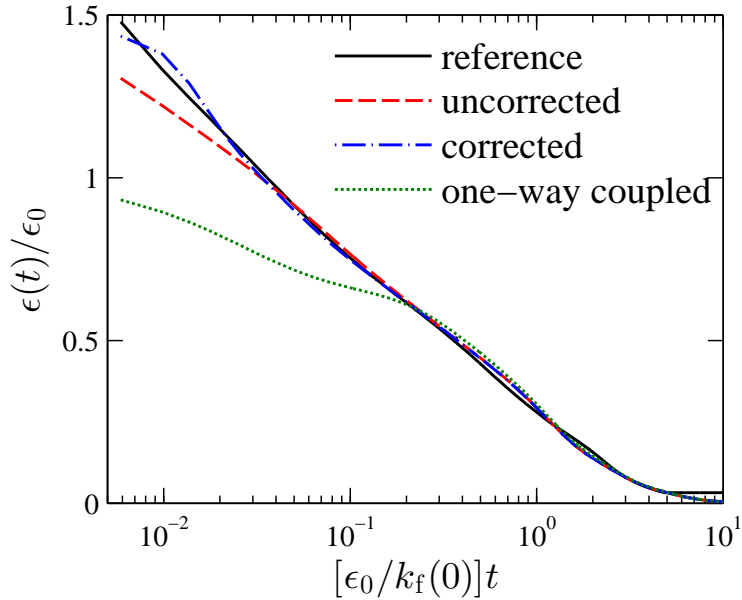}
\caption{The rate of energy dissipation in particle-laden decaying homogeneous turbulence as a function of time normalized by $\epsilon_0$, which is the dissipation rate of the unladen flow at $t=0$.
More details are provided in the caption of Figure \ref{fig:Kp} (color online.)}
\label{fig:epsilon}
\end{center}
\end{figure}

At 1.8\%, the mass fraction of particles is too small to contrast the effect of the correction scheme on $k\f$. 
Thus, in the interest of brevity, we forgo reporting $k\f$ for various simulations as all curves lie on top of each other.
However, the particle model has an effect on the rate of energy dissipation $\epsilon(t)$, computed using
\begin{equation}
\epsilon(t) = -\frac{{\rm d}}{{\rm d}t}\left[ (1-\phi_{\rm m}) k\f(t) + \phi_{\rm m} k\p(t) \right],
\label{epsilon-FK}
\end{equation}
where $\phi_{\rm m}$ is the mass-fraction ratio. 
Alternatively, by following the derivation in \cite{subramaniam_et_al_2014}, $\epsilon$ can be computed using
\begin{equation}
\epsilon(t) = \frac{1}{\rho\f V}\left[ -\int_V \bl u \cdot \nabla\cdot \left(\mu \nabla \bl u\right) {\rm d} V + \sum_{i=1}^{N\p} \bl F_i(t) \cdot \left(\bl u_{{\rm d},i} - \bl u_{{\rm p},i}\right) \right],
\label{epsilon}
\end{equation}
where $\bl F_i(t)$ and $\bl u_{{\rm d},i}(t)=\bl u(\bl x_{{\rm p},i},t)$ are the two-way coupling force (modeled using Eq. \eqref{F_exp}) and the disturbed fluid velocity computed at the location of $i^{\rm th}$ particle, respectively.
Although Eq. \eqref{epsilon} is more insightful for analysis of the dissipations curves in Figure \ref{fig:epsilon}, we used Eq. \eqref{epsilon-FK} to compute $\epsilon$ in practice, which permits us to make an apple-to-apple comparison against particle-resolved simulation.
In this case study, since the initially stagnant particles are suddenly introduced in the flow, there is a very short transient period at which a boundary layer develops around particles. 
As elaborated in \ref{sec:appC}, within this short transient period $\epsilon$ varies significantly as the local velocity gradients change rapidly. 
The time scale associated with this transient period is too short to be resolved by the point-particle simulation, which runs at much larger time step size than the particle-resolved simulation.
Since nonlinearly varying $\epsilon$ is computed in a discrete setting by averaging over a time step, averaging it at two different time step size could yield drastically different results. 
To circumvent this issue, we employ Eq. \eqref{epsilon-FK} to compute $\epsilon$ using the following procedure.
$k\f$ and $k\p$ from the particle-resolved simulation are interpolated to the points in time at which point-particle $k\f$ and $k\p$ are available, i.e., at $\Delta t, 2\Delta t, \cdots$ with $\Delta t$ being the time step size of the point-particle simulation. 
Then the dissipation-rate is computed at $(i+1/2)\Delta t$ based on the difference between the total kinetic energy at $(i+1)\Delta t$ and $i\Delta t$. 
Following this procedure, the reported dissipation-rate in Figure \ref{fig:epsilon} correspond to the same interval, allowing for a one-to-one comparison between particle-resolved and point-particle simulations. 

Figure \ref{fig:epsilon} shows that $\epsilon$ is most sensitive to the point-particle model at short time when $\|\bl u\p \| \approx 0$ and $\|\bl u_{\rm d} - \bl u\p\|$ is relatively large.
This observation can be explained based on Eq. \eqref{epsilon}. 
The first and second terms in this equation represent the fluid dissipation resolved on the grid and those at the scale of the particle that is modeled by the point particle-approach, respectively.
The second term that needs not to be included in the particle-resolved simulation is included in the point-particle simulations to account for the unresolved scales. 
The difference between various point-particle curves in Figure \ref{fig:epsilon} is primarily attributed to the form of the second term in each particle model. 
In the case of the one-way coupled simulation, $\bl F_i\cdot \bl u_{{\rm d},i}$ terms are entirely neglected (hence $\epsilon(0) \ne \epsilon_0$ in Figure \ref{fig:epsilon}), whereas, for the two-way coupled simulations, it is modeled.
What differentiates the corrected and uncorrected schemes is how $\bl F_i$ is modeled. 
Neglecting the finite Reynolds number effects, $\bl F$ must be proportional to $\bl u\f - \bl u\p$ according to the Stokes drag formula. 
The uncorrected scheme assumes $\bl u\f = \bl u_{\rm d}$, a quantity that lies between the undisturbed fluid velocity and particle velocity.
Consequently, the scheme without correction underestimates $\bl F_i$ and subsequently $\epsilon$, yet provides a better approximation than the one-way coupled scheme that completely neglects a part of the second term in Eq. \eqref{epsilon}.
The present correction scheme, on the other hand, provides a much better estimate of $\bl F_i$, yielding a better estimate of the model form dissipation even at the very short time (Figure \ref{fig:epsilon}).

\section{Discussion and Conclusion}
Two-way coupled Euler-Lagrange simulations can suffer from significant errors introduced by the wrong estimation of the undisturbed fluid velocity.
Employing the velocity of the fluid at the location of the particle as the undisturbed fluid velocity can become inaccurate as the disturbance created by the particle becomes large.
We argued that the velocity disturbance created by a particle in the discrete setting can be modeled as the response of a solid object to the two-way coupling forces. 
That solid object is the computational cell that hosts the particle.
This analogy allowed us to write the equation of the motion of a computational cell as a solid object dragged by the two-way coupling forces in the fluid. 
The outcome of this procedure was an estimation of the velocity of the computational cell that was used in correcting the disturbed velocity of the fluid at the location of the particle, obtaining a correction scheme that yielded the undisturbed fluid velocity. 
The final formulation is general in scope as it can be applied to various grid geometry and particle and flow conditions. 
Tests of the proposed correction scheme using isotropic and anisotropic grids with various sizes for a single particle settling with different Reynolds and Stokes numbers. 
The present correction scheme was able to predict the settling velocity with a few percent error, lowering the error associated with the uncorrected simulations by one to two orders of magnitude. 
A similar calculation with two nearby particles settling side-by-side showed that the present correction scheme could correctly predict the modification in the settling velocity due to the screening effect.
Excellent agreement was also achieved between the corrected point-particle simulation and the particle-resolved simulation of decaying turbulence in terms of flow energetics.
The cases attested to the accuracy of our correction scheme in estimating the undisturbed velocity in two-way coupled point-particle simulations for a wide range of physical and numerical conditions.

An empirical equation was employed for estimation of the drag force on the non-spherical computational cell (Eq. \eqref{kc}). 
The implementation of this relationship is simple in practice when compared against a two-dimensional tabulation of the drag coefficient as a function of $a^{(2)}/a^{(1)}$ and $a^{(3)}/a^{(1)}$. 
The cost of this convenience is an error in estimation of $K\c^{(i)}$, depicted by the distance of the points from the bisector in Figure \ref{fig:Kc}.
This wrong estimation was manifested in cases N06, N09, and N10, where a larger relative error was observed for grids with higher aspect ratio.
Further refinement of Eq. \eqref{kc} or its tabulation is required to cover a broader range of aspect ratios while maintaining a reasonable accuracy.
Furthermore, problems dealing with complex geometries are generally discretized using unstructured grids (e.g., tetrahedral elements). 
We kept the formulation in Section \ref{sec:method} general, such that it can be adopted for any arbitrary cell geometry.
What changes in this formulation for an arbitrary element is the drag coefficient $K_{\rm c}^{(i)}$, which must be extracted empirically or calibrated numerically. 
Developing a similar empirical equation for arbitrary-shaped elements or testing available equations from \cite{leith1987drag} for such cases remains to be explored in future studies.

The simulations involving a pair of particles attested to the potential of the present correction scheme in capturing the interaction of nearby particles with a reasonable accuracy.
According to cases T01 to T04, the ideal grid size for most accurate prediction of such interactions is of the same order as the particle diameter. 
For a grid much larger than the particle, the resolution is not sufficient to resolve the disturbance field, whereas, for a grid much smaller than the particle, the error associated with a single particle calculation hinders the overall accuracy of the scheme. 
In the latter limit, developing a hybrid point-particle and immersed-boundary approach in the future where the two-way coupling forces are distributed among cells that overlay with the particle should result in a more computationally-consistent scheme with potentially better convergence characteristics.

The decaying turbulence case showed that the present scheme is promising in simulating more complex flows, which otherwise would require the large computational demand of a particle-resolved simulation with over three orders of magnitude higher computational cost.
Although a remarkable agreement with the particle-resolved simulation was observed for that case, we can not conclude that the present scheme can fully capture particle-particle interaction in a complex turbulent flow. 
That is because the volume fraction of the studied case was too low for particle-particle interactions to have a detectable effect on the results. 
Therefore, what remains to be done in the future is comparing the present point-particle scheme to a particle-resolved simulation at a much larger volume fraction to verify its validity and accuracy in capturing particle-particle interactions.

The necessity of having a correction scheme for point-particle simulations is not limited to the momentum equation. 
For instance, the problem of heat transfer from a particle to its surrounding flow is modeled based on the undisturbed fluid temperature, a quantity that is often replaced by the fluid temperature at the location of the particle.
This naive substitution introduces an error similar to the one discussed in this article. 
Extending the present scheme to such scenarios requires reformulation of Eq. \eqref{uc_ge} to express the temperature of the computational cell as the temperature of a solid object surrounded by the fluid.
Various elements of $K_{\rm t}^{(i)}$ would be modified accordingly for the thermal problem.
The correction for the drag coefficient (Eq. \eqref{kc}) will be substituted by a correction for the thermal resistance of an arbitrary-shaped cell. 
The structure of the correction for the interpolation (Eq. \eqref{kp}) remains the same with a similar relationship to Eq. \eqref{alpha} that is solely a function of $1/r_{jk}$.
A similar correction for the finite Reynolds will be needed and the correction for the exposure time remains the same. 
Thus, the present framework can be expanded to obtain a correction scheme for other partial differential equations that are coupled to a point-particles model. 

\section*{Acknowledgments} 
We thank Shankar Subramaniam and Mohammad Mehrabadi for sharing their particle-resolved simulation results and Ali Mani and Shima Alizadeh for fruitful discussions. 
This work was funded by the United States Department of Energy's (DoE) National Nuclear Security Administration (NNSA) under the Predictive Science Academic Alliance Program II (PSAAP II) at Stanford University.
We acknowledge the use of computational hours on the Certainty cluster at Stanford University, where all these tools have been developed. 

\appendix
\section{A simplified algebraic correction scheme} \label{sec:simplified}
The cost associated with the scheme described in Section \ref{sec:comp} is of the same order as the original particle tracking problem. 
It requires three additional numbers, i.e. $\bl u\c$, to be stored per particles and to be integrated in time.
Therefore, with regard to both processing and memory requirements, it approximately doubles the needed computational resources associated with the particle-tracking problem. 
If the added memory requirement is a limiting factor or one wishes to keep the structure of the code unchanged, it is possible to solve an algebraic equation instead of integrating Eq. \eqref{uc_ge} numerically in time. 
That is by neglecting the time dependent term in Eq. \eqref{uc_ge} and directly computing
\begin{equation}
u\c^{(i)} \approx -\frac{F^{(i)}}{3\pi\mu d\c K_{\rm t}^{(i)}}.
\label{uc_sim}
\end{equation}
This equation can be either solved iteratively along with steps 2, 3, 8, and 10 above or combined with the equation of the motion of particle to find a stand-alone equation.
For instance, if $\bl F$ is governed by Eq. \eqref{stokes} and $\rm Re\c \ll 1$ such that $C_{\rm r}$ can be approximated as 1 ($K\c^{(i)}$, $K\p^{(i)}$, and $C_{\rm t}^{(i)}$ are all independent of $\bl u\c$), one can combine Eqs. \eqref{particle_eq}, \eqref{stokes}, \eqref{ufe_def}, and \eqref{uc_sim} to show
\begin{equation}
F_{\rm d}^{(i)} = 3\pi\mu d\p K^{(i)} \left(u_{\rm d}^{(i)} - u\p^{(i)}\right)
\label{uc_one_line}
\end{equation}
with 
\begin{equation}
K^{(i)} = \left(1-\frac{d\p}{d\c K_{\rm t}^{(i)}}\right)^{-1}.
\label{K_def}
\end{equation}
With these simplifications, the entire algorithm reduces to the calculation of $K^{(i)}$, which must be incorporated into the Stokes drag that otherwise is computed based on the disturbed fluid velocity. 
In other words, the above algorithm can be restructured as going over steps 5 to 10 in Section \ref{sec:comp} to compute $K^{(i)}_{\rm t}$, then using Eqs. \eqref{K_def} and \eqref{uc_one_line} to compute $\bl F$, and finally computing $\dot {\bl u}\p$ using Eq. \eqref{particle_eq}.
It is even possible to substitute $K_{\rm t}^{(i)}$ with a constant and refine it later on by accounting for $K\c^{(i)}$, $K\p^{(i)}$, $C_{\rm r}$, and $C_{\rm t}^{(i)}$ in a step-by-step fashion.
Thus, the crudest version of this correction scheme for an isotropic grid translates to simply dividing the Stokes drag computed based on the disturbed velocity by $1 - 0.75 \Lambda$ with $\Lambda$ denoting the ratio of the particle diameter to the grid size. 

The above simplification, which obviates the need for calculation and storage of $\bl u\c$, comes at a cost. 
As implied by Eq. \eqref{K_def}, $K^{(i)} \to \infty$ when $d\c K_{\rm t}^{(i)} \to d\p$, i.e. the simplified algorithm diverges if the particle and the computational cell have similar sizes.
$K_{\rm t}^{(i)}$ is generally larger than 0.5, hence it is only safe to use Eq. \eqref{uc_one_line} for correction when the grid size is at least twice as large as the particle.
In such scenarios ($\Lambda \le 1$), the accuracy remains roughly similar to the full algorithm that involves time integration of $\dot {\bl u}\c$.
Note that the full algorithm described in Section \ref{sec:comp} does not encounter the above singularity and is not limited to relatively small particle sizes.

\section{The history effect} \label{app:history}

In general, the effect of the history term on the solution is non-trivial.
In what follows, we estimate its effect for a particular scenario in which a particle accelerates from a stationary condition and reaches a terminal velocity.
This correction would be particularly useful in studying systems similar to those considered under Section \ref{sec:result} with particles settling under gravity.
In each of these scenarios, the velocity of the particle in the transient period follows a trend that is approximately an exponential function with a time constant $\tau\p$ (Figure \ref{fig:up}).
Similar to the particle, the computational cell also transients from a stationary condition to reach a terminal velocity with a time constant $\tau\c^{(i)}$ according to Eq. \eqref{transient_sol}.
Note that the history effect alters the general form of the solution such that the exponential behavior in Eq. \eqref{transient_sol} may no longer hold.
However, this exponential variation is used here as an approximation to delineate the effect of the history term on $\bl u\c$.

To include the history effect, we consider
\begin{equation}
\frac{3}{2}m\c \dot u\c^{(i)} = -3\pi\mu d\c K_{\rm t}^{(i)} \left( u\c^{(i)} + \frac{1}{2} d\c \int_{-\infty}^{t}[\pi\nu(t-t^\prime)]^{-\frac{1}{2}}\dot u\c^{(i)}(t^\prime) {\rm d}t^\prime\right) - F^{(i)},
\label{uc_wh}
\end{equation}
instead of Eq. \eqref{uc_ge}, in which the second term in the right-hand side is the history term. 
In the transient period, $\dot {\bl u}\c$ reaches its maximum.
Provided the proportionality of the history term to $\dot {\bl u}\c$, we focus on this period in our calculations.
Therefore, we estimate in Eq. \eqref{uc_wh} assuming $\bl u\c$ changes according to Eq. \eqref{transient_sol}.

\begin{figure}
\begin{center}
\includegraphics[width=0.6\textwidth]{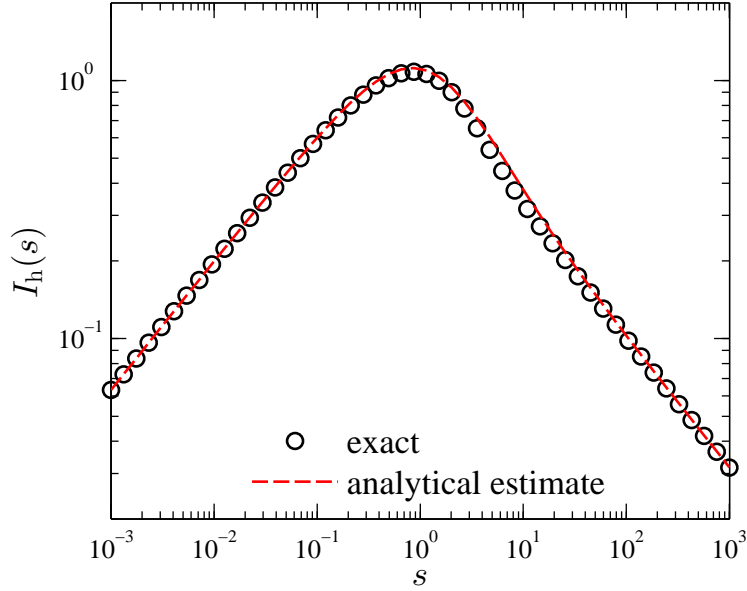}
\caption{Direct numerical calculation (black circles) versus analytical estimate (dashed red line) of the history term integral in Eq. \eqref{Ih}.}
\label{fig:Ih}
\end{center}
\end{figure}

For scenarios under consideration, the particle is introduced at $t=0$, and the disturbance field is zero for $t<0$.
Thus, taking $t_0=0$ in Eq. \eqref{transient_sol}, we have $\bl u\c = 0$ for $t < 0$ and 
\begin{equation}
\dot u\c^{(i)}(t) = \frac{U\c^{(i)}}{\tau\c^{(i)}}\exp\left(-\frac{t}{\tau\c^{(i)}}\right)
\label{udotc}
\end{equation}
for $t > 0$.
Using Eq. \eqref{udotc} for $\dot u\c^{(i)}$ and substituting it in the second term of Eq. \eqref{uc_wh} yields
\begin{equation}
\frac{1}{2}d_c \int_0^t [\pi\nu(t-t^\prime)]^{-\frac{1}{2}} \dot u\c^{(i)}(t^\prime)\; {\rm d}t^\prime = U\c^{(i)} \left(\frac{3}{\pi}K\c^{(i)}\right)^{1/2} \int_0^{t/\tau\c^{(i)}}{ \left[(t-t^\prime)/\tau\c^{(i)} \right]^{-1/2} \exp{\left(-t^\prime/\tau\c^{(i)}\right)}\; {\rm d}\left(t^\prime/\tau\c^{(i)} \right) }.
\label{hist_expanded}
\end{equation}
The integral in Eq. \eqref{hist_expanded} can be calculated numerically for any given $t/\tau\c^{(i)}$. 
Further examination of this integral (Figure \ref{fig:Ih}) shows that it can be estimated using 
\begin{equation}
I_{\rm h}(s) = \frac{(4+s)\sqrt s}{2+1.5s+s^2} \approx \int_0^{s} \left[(s-s^\prime)\right]^{-1/2} e^{-s^\prime}\; {\rm d}s^\prime.
\label{Ih}
\end{equation}

Using Eq. \eqref{Ih} as an analytical estimate of the history term integral allows writing Eq. \eqref{uc_wh} as  
\begin{equation}
\frac{3}{2}m\c \dot u\c^{(i)} = -3\pi\mu d\c K_{\rm t}^{(i)} \left[ u\c^{(i)} + U\c^{(i)} \left(\frac{3}{\pi}K\c^{(i)}\right)^{1/2} I_{\rm h}\left(\frac{t}{\tau\c^{(i)}}\right) \right] - F^{(i)}.
\label{uc_w_hist}
\end{equation}
In order to obtain an estimate of $U\c^{(i)}$, we consider the limit $t\to\infty$ at which $\bl u\c = \bl U\c$ and $\dot{\bl u}\c = 0$. 
Thus, from Eq. \eqref{uc_w_hist}
\begin{equation}
 U\c^{(i)} = -\left[1 + \left(\frac{3}{\pi}K\c^{(i)}\right)^{1/2} I_{\rm h}\left(\frac{t}{\tau\c^{(i)}} \right) \right]^{-1} \frac{F^{(i)}}{3\pi\mu d\c K_{\rm t}^{(i)}}.
\label{uc_w_hist_terminal}
\end{equation}
Finally, combining Eqs. \eqref{uc_w_hist} and \eqref{uc_w_hist_terminal} yields
\begin{equation}
\frac{3}{2}m\c \dot u\c^{(i)} = -3\pi\mu d\c K_{\rm t}^{(i)} u\c^{(i)} - \frac{F^{(i)}}{C_{\rm h}^{(i)} },
\label{hist}
\end{equation}
where 
\begin{equation}
C_{\rm h}^{(i)} = 1 + \left(\frac{3}{\pi}K\c^{(i)}\right)^{1/2} I_{\rm h}\left(\frac{t}{\tau\c^{(i)}}\right).
\label{Ch}
\end{equation}
The above equation expresses how the correction for the history term should be incorporated in the framework outlined in Section \ref{sec:comp}. 
In the modified framework, step 11 is substituted by computing $C_{\rm h}^{(i)}$ from Eq. \eqref{Ch} and using it to integrate Eq. \eqref{hist} rather than Eq. \eqref{uc_ge}. 

Contrary to all other corrections, the history correction acts on the force rather than the drag coefficient. 
As we showed in Section \ref{sec:method}, the geometry of the computational cell, the interpolation scheme, and the finite Reynolds number and finite exposure effects all modified the drag coefficient through $K_{\rm t}^{(i)}$. 
Despite the fact that the history term also appears in the same group as the drag force in Eq. \eqref{uc_wh}, its effect is captured through $C_{\rm h}^{(i)}$ by modifying the two-way coupling force $\bl F$. 

\begin{figure}
\begin{center}
\includegraphics[width=0.6\textwidth]{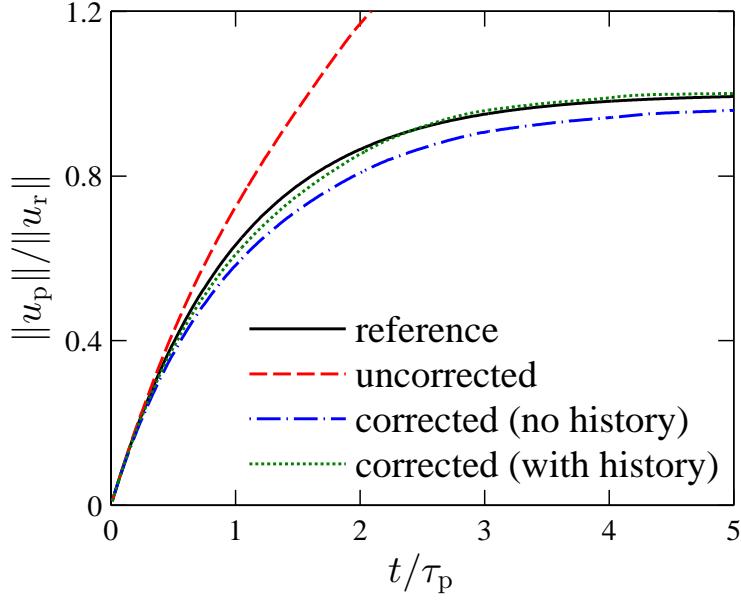}
\caption{The effect of correction for the history effect on the settling velocity of a particle in the transient period.
All the curves except dotted green, which is obtained by including the correction for the history term, are replicated from Figure \ref{fig:up}.
The history correction is employed only for the computational cell and neglected in the particle equation of motion in all cases.}
\label{fig:up_w_hist}
\end{center}
\end{figure}

The correction scheme that was used in Section \ref{sec:result} can be considered to be governed by Eq. \eqref{hist} when $C_{\rm h}^{(i)}=1$. 
Accounting for $C_{\rm h}^{(i)}$ via Eq. \eqref{Ch}, however, does not affect the particle settling terminal velocity.
Thus, all the earlier reported errors, which are computed based on $\bl u\p$ at $t\to\infty$, are independent of the history correction. 
At this limit, $I_{\rm h} \to 0$ and $C_{\rm h}^{(i)} \to 1$, reducing Eq. \eqref{hist} to Eq. \eqref{uc_ge}. 
Employing Eq. \eqref{hist} for the correction scheme, however, has an effect on the transient period at which the particle experiences high acceleration (Figure \ref{fig:up_w_hist}).

The importance of the history term correction depends on the value of $I_{\rm h}$.
At its maximum, which occurs at $t \approx 0.85 \tau\c^{(i)}$, $I_{\rm h} \approx 1.1$.
Since $\left(3K\c^{(i)}/\pi\right)^{1/2} = \mathcal O(1)$, $C_{\rm h}^{(i)}$ effect is limited in reducing the force at most by a factor of two.
Smaller $\bl F$, decreases $\bl u\c$, thus increasing the settling velocity of the particle through a larger $\bl u\f$ (Figure \ref{fig:up_w_hist}).
On the other extent at which $I_{\rm h}$ is minimum (either $t \to 0$ or $t \to \infty$), the effect of the history term vanishes since $C_{\rm h} \to 0$.
In conclusion, the correction for the history only increases the settling velocity of a particle (by reducing its drag), and its effect is limited to $t = \mathcal O(\tau\c^{(i)})$ at which particle acceleration is the largest.

\section{The early-time dissipation-rate in particle-laden isotropic turbulence} \label{sec:appC}

In the results section, the dissipation rate at early times of the particle-resolved calculation was averaged to a time scale equal to that of the point-particle schemes before comparing the two. The dissipation rate at very early times is not resolved by the particle-resolved method. In both particle-resolved and point-particle simulations, the initial condition of the particles is at rest relative to the fluid. Though such an initial condition cannot be physically realized in practice, it may be analyzed theoretically. Neglecting the curvature of particle surfaces, the flow near each particle surface at early times can be approximated by Stokes' first problem. The equation for the flow \cite{white-2006} is 
\begin{equation}
u = U_{\rm o}{\rm erf}(r/ 2 \sqrt[]{\nu t}),
\end{equation}
where $r$ is the radial coordinate of the particle. The instantaneous dissipation rate is 
\begin{equation}
\varepsilon(r,t) = \nu (\frac{\partial u}{\partial y})^{2} = U_{\rm o}^{2}(\pi\nu t)^{-1}{\rm exp}(-r^2/2\nu t).
\label{Stfdisp}
\end{equation}

The mean dissipation rate in the fluid volume can be found by integrating Eq.~\ref{Stfdisp} over the radial coordinate:
\begin{equation}
\varepsilon(t) = \lim_{\beta \rightarrow\infty} U_{\rm o}^{2}(\pi\nu t)^{-1}\int_{0}^{\beta} {\rm exp}(-r^2/2\nu t) dr = \lim_{\beta \rightarrow\infty} \frac{U_{o}^{2}\sqrt{\nu}}{\sqrt{2\pi t}\beta}.
\label{Stfdispavg}
\end{equation}

The final term in Eq.~\ref{Stfdispavg} is an indeterminate form at $t = 0$ and requires interpretation. On the one hand, the diffusion solution predicts a disturbance arbitrarily far from the particle boundary at an arbitrarily early time. The disturbance, however, is limited by the sound speed of the fluid. Alternatively, we may say that in practice, any physical domain is of bounded size. Both explanations render the limit $\beta$ to some real number less than infinity. The average dissipation at $t = 0$ for any finite domain is therefore infinite. In other words, it takes an infinite amount of work to accelerate the particle from a state of rest to a new fluid velocity equal to the undisturbed fluid velocity at the location of the particle at $t = 0^{-}$. Though dissipation is singular at the initial condition, it's integral in time is regular, so that the change in kinetic energy of the fluid is finite. This explanation is provided to demonstrate that the particle-resolved dissipation at very early times is under-resolved. The dissipation as computed in the particle-resolved simulation begins at the nominal dissipation rate of the initial condition. Following, there is a quick development of boundary layers around each particle. This corresponds to a local maximum in the dissipation rate computed by the particle-resolved calculation between a non-dimensional time of $10^{-3}-10^{-2}$. Following this peak, the boundary layers have been resolved on the fluid grid and the dissipation rate as computed in the particle-resolved simulation maybe considered physical. This is approximately the non-dimensional time where we compare the dissipation rate computed in the particle-resolved and the point-particle simulations.


\begin{figure}
\begin{center}
\includegraphics[width=0.6\textwidth]{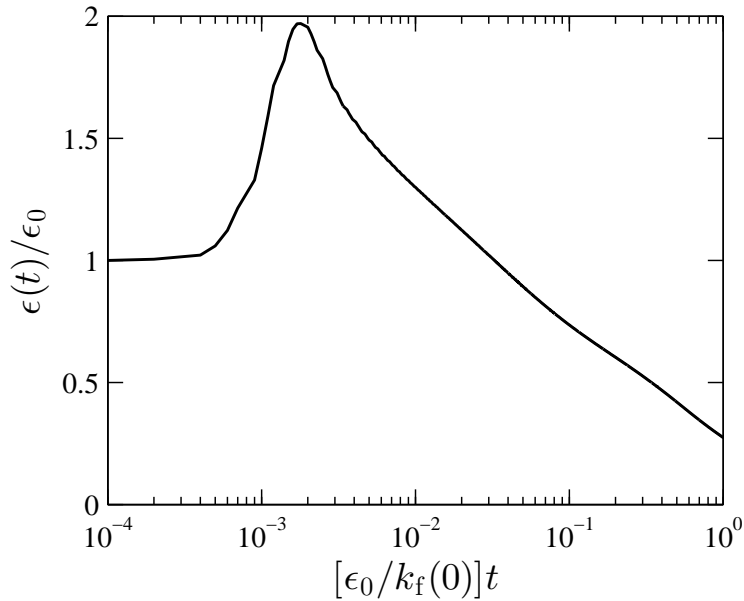}
\caption{The rate of energy dissipation in particle-laden decaying homogeneous turbulence as a function of time obtained from particle-resolved simulation.}
\label{fig:epsilon-PR}
\end{center}
\end{figure}

\def\bibsection{
\end{document}